\documentclass[useAMS,usenatbib]{mn2e} 
\usepackage{graphicx}

\newcommand{\apj}{ApJ}

\newcommand{\aap}{A\&A}

\newcommand{\e}[1]{$\times10^{#1}$}
\newcommand{\gppr}{\stackrel{>}{\scriptstyle \sim}}
\newcommand{\gappr}{\raisebox{-0.4ex}{$\gppr$}}

\newcommand{\Teff}{\mbox{$T_{\mathrm{eff}}$}}

\newcommand{\Porb}{\mbox{$P_{\mathrm{orb}}$}}

\newcommand{\Lines}[3]{\Ion{#1}{#2}\,$\lambda\lambda$\,#3}
\newcommand{\Ion}[2]{#1{\,\scriptsize #2}}
\newcommand{\id}{\mbox{$\mathrm{d^{-1}}$}}
\newcommand{\kms}{\mbox{$\mathrm{km\,s^{-1}}$}}

\title[Post Common  Envelope Binaries from  SDSS]{Post Common Envelope
  Binaries from SDSS - III. Seven new orbital periods}
\author[A.  Rebassa-Mansergas et al.]{A.  Rebassa-Mansergas$^1$, B.
  T.  G\"ansicke$^1$, M.R.  Schreiber$^2$,
  J. Southworth$^1$,\newauthor A.D. Schwope$^3$, A.  Nebot
  Gomez-Moran$^3$, A.  Aungwerojwit$^{4,1}$, P.
  Rodr\'{\i}guez-Gil$^{5,6}$,\newauthor V.  Karamanavis$^7$, M.
  Krumpe$^3$, E.  Tremou$^7$, R.  Schwarz$^3$, A.  Staude$^3$,
  J. Vogel$^3$\\
$^{1}$ Department of Physics, University of Warwick, Coventry CV4 7AL, UK \\
$^{2}$ Departamento de F\'\i sica y Astronom\'\i a, Universidad de Valpara\'\i so, 
Avenida Gran Bretana 1111, Valpara\'\i so, Chile \\
$^{3}$ Astrophysikalisches Inst. Potsdam, An der Sternwarte 16, 14482, Potsdam, Germany\\
$^{4}$ Department of Physics, Faculty of Science, Naresuan University, Phitsanulok,
65000, Thailand\\
$^{5}$ Isaac Newton Group of Telescopes, Apdo. de Correos 321, E-38700, Santa Cruz
de La Palma, Spain\\
$^{6}$ Instituto de Astrof\'\i sica de Canarias, V\'\i a L\'actea, s/n, La Laguna, 
E-38205, Tenerife, Spain\\
$^{7}$ Dept. of Physics, Sect. of Astrophysics, Astronomy \& Mechanics,
Univ. of Thessaloniki, 541 24 Thessaloniki, Greece 
}

\begin{document}
\date{Accepted 2008. Received 2008; in original form 2008}
\pagerange{\pageref{firstpage}--\pageref{lastpage}} \pubyear{2008}
\maketitle

\begin{abstract}
We  present follow-up spectroscopy  and photometry  of 11  post common
envelope  binary  (PCEB)  candidates  identified from  multiple  Sloan
Digital  Sky Survey (SDSS)  spectroscopy in  an earlier  paper. Radial
velocity   measurements   using   the   \Lines{Na}{I}{8183.27,8194.81}
absorption  doublet  were performed  for  nine  of  these systems  and
provided  measurements of  six orbital  periods in  the  range $\Porb=
2.7-17.4$\,h.  Three  PCEB candidates did not  show significant radial
velocity  variations  in  the  follow-up  data,  and  we  discuss  the
implications for the use of SDSS spectroscopy alone to identify PCEBs.
Differential  photometry confirmed  one of  our  spectroscopic orbital
periods  and  provided  one  additional \Porb\,  measurement.   Binary
parameters  are estimated  for the  seven  objects for  which we  have
measured the orbital  period and the radial velocity  amplitude of the
low-mass companion star, $K_\mathrm{sec}$.   So far, we have published
nine SDSS PCEBs orbital periods,  all of them $\Porb<1$\,d. We perform
Monte-Carlo simulations  and show that $3\sigma$  SDSS radial velocity
variations  should still  be  detectable for  systems  in the  orbital
period  range  of  $\Porb\sim1-10$\,days.  Consequently,  our  results
suggest  that   the  number   of  PCEBs  decreases   considerably  for
$\Porb>1$\,day, and that during  the common envelope phase the orbital
energy of the binary star is maybe less efficiently used to expell the
envelope than frequently assumed.

\end{abstract}


\label{firstpage}

\section{Introduction}

Most of the stellar sources are  formed as parts of binary or multiple
systems. Therefore  the study of  binary star evolution  represents an
important part  of studying stellar  evolution. While the  majority of
the wide  main sequence binaries evolve  as if they  were single stars
and  never interact,  a small  fraction are  believed to  undergo mass
transfer  interactions.   Once the  more  massive  main sequence  star
becomes  a   red  giant  it  eventually  overfills   its  Roche  lobe.
Dynamically  unstable  mass  transfer  exceeding the  Eddington  limit
ensues onto the companion  star, which consequently also overfills its
own Roche  lobe.  The  two stars then  orbit inside a  common envelope
(CE,  e.g.   \citealt{paczynski76-1, livio+soker88-1,  iben+livio93-1,
taam+ricker06-1,  webbink07-1}),  and  friction inside  this  envelope
causes a rapid  decrease of the binary separation.  Orbital energy and
angular momentum are  extracted from the binary orbit  and lead to the
ejection  of the  envelope,  exposing a  post  common envelope  binary
(PCEB).  After the envelope is expelled PCEBs keep on evolving towards
shorter   orbital   periods   through   angular  momentum   loss   via
gravitational  radiation  and, for  companion  stars  above the  fully
convective mass limit,  by magnetic wind braking. In  binaries that do
not undergo a CE phase both components evolve almost like single stars
and keep  their wide binary  separations.  A predicted  consequence is
hence  a  strongly  bi-modal  binary  separation  and  orbital  period
distribution   among   post-main   sequence   binaries,   with   PCEBs
concentrated at  short orbital periods, and non-PCEBs  at long orbital
periods.

\begin{table*}
\label{t-logobs}
\caption{Log of  the observations. Included are the  target names, the
  SDSS  $ugriz$ psf-magnitudes,  the period  of the  observations, the
  telescope/instrument  setup, the  exposure time,  and the  number of
  exposures.}  
\setlength{\tabcolsep}{1.0ex}
\begin{tabular}{lcccccccccrlr}
\hline
\hline
\textbf{Spectroscopy}    &                   &       &        &             &           &     \\
SDSS\,J  & 
$u$ & $g$ & $r$ & $i$ & $z$ & Dates & Telesc. & Spectr. & Grating/Grism & Exp.[s] & \# spec. \\
\hline
005245.11--005337.2 &  20.47 & 19.86 & 19.15 & 17.97 & 17.22 & 16/08/07-20/08/07 & NTT        & EMMI   & Grat\#7     & 1200      & 21  \\     
024642.55+004137.2  &  19.99 & 19.23 & 18.42 & 17.29 & 16.60 & 05/10/07-09/10/07 & NTT        & EMMI   & Grat\#7     & 650-800   & 19  \\ 
030904.82--010100.8 &  20.77 & 20.24 & 19.50 & 18.43 & 17.77 & 07/10/07-12/10/07 & NTT        & EMMI   & Grat\#7     & 1500      & 4   \\
031404.98--011136.6 &  20.78 & 19.88 & 19.03 & 17.76 & 16.98 & 02/10/07-03/10/07 & M-Baade    & IMACS  & 600 line/mm & 600-900   & 12  \\
113800.35--001144.4 &  19.14 & 18.86 & 18.87 & 18.15 & 17.53 & 16/08/06-23/03/07 & VLT ($sm$) & FORS2  & 1028z       & 900       & 2   \\ 
                    &        &       &       &       &       & 18/06/07-23/06/07 & WHT        & ISIS   & 158R        & 1500      & 1   \\
                    &        &       &       &       &       & 17/05/07-20/05/07 & M-Clay     & LDSS-3 & VPH-red     & 500-600   & 5   \\
115156.94--000725.4 &  18.64 & 18.12 & 18.14 & 17.82 & 17.31 & 17/05/07-20/05/07 & M-Clay     & LDSS-3 & VPH-red     & 450-600   & 17  \\ 
152933.25+002031.2  &  18.69 & 18.20 & 18.33 & 17.98 & 17.48 & 18/05/07-20/05/07 & M-Clay     & LDSS-3 & VPH-red     & 500-1000  & 18  \\ 
172406.14+562003.0  &  15.83 & 16.03 & 16.42 & 16.42 & 16.52 & 25/09/00-29/03/01 & SDSS       &        &             & 900       & 23  \\
224139.02+002710.9  &  19.62 & 18.82 & 18.40 & 17.33 & 16.58 & 02/10/07-03/10/07 & M-Baade    & IMACS  & 600 line/mm & 600       & 2   \\
                    &        &       &       &       &       & 07/10/07-12/10/07 & NTT        & EMMI   & Grat\#7     & 750-800   & 3   \\
233928.35--002040.0 &  20.40 & 19.68 & 19.15 & 18.07 & 17.36 & 07/10/07-12/10/07 & NTT        & EMMI   & Grat\#7     & 1400-1700 & 15  \\ 
\hline
\hline
\textbf{Photometry}      &                   &       &        &             &           &     \\
SDSS\,J  & $u$ & $g$ & $r$ & $i$ & $z$ & Dates & Telesc. &  & Filter band   &  Exp.[s]   & \# hrs  \\
\hline
031404.98--011136.6 &  20.78 & 19.88 & 19.03 & 17.76 & 16.98 & 19/09/06-20/09/06 & CA\,2.2\,m  & & clear & 35-60  & 10.3 \\ 
082022.02+431411.0  &  15.92 & 15.85 & 16.11 & 15.83 & 15.38 & 21/11/06-24/11/06 & Kryoneri\,1.2\,m & & $R$ & 60   & 8.2 \\
172406.14+562003.0  &  15.83 & 16.03 & 16.42 & 16.42 & 16.52 & 04/08/06-10/08/06 & IAC80       & & $I$   & 80-180 & 12.2 \\
                    &        &       &       &       &       & 25/03/06-13/09/06 & AIP 70 cm   & & $R$   & 60-90  & 55.6 \\
\hline
\end{tabular}
\begin{minipage}{\textwidth}
  Notes: M-Baade  and M-Clay refer  to the two Magellan  telescopes at
  Las Campanas observatory. We use $sm$ to indicate that the data were
  taken in service  mode. CA\,2.2 is the 2.2  metre telescope at Calar
  Alto observatory.
\end{minipage}
\end{table*}

The scenario outlined  above is thought to be  a fundamental formation
channel  for a  wide range  of astronomical  objects such  as low-mass
X-ray binaries, double degenerate white dwarfs, neutron star binaries,
cataclysmic  variables and  super-soft X-ray  sources.  Some  of these
objects  will eventually  end their  lives as  type Ia  supernovae and
short gamma-ray bursts, which are of great importance for cosmological
studies.

While the  basic scenario of  CE evolution has  been known for  a long
time \citep[e.g.][]{paczynski76-1}, the  physical details involved are
complex and still poorly  understood. As a result, full hydrodynamical
models for the  CE phase have been calculated only  for a small number
of cases \citep{sandquistetal00-1,ricker+taam08-1}. Similarly, orbital
evolution  through  magnetic  braking  is not  well  understood,  with
different prescriptions  in angular momentum loss  differing by orders
of   magnitudes  \citep[e.g.][]{verbunt+zwaan81-1,  andronovetal03-1}.
Consequently, current binary population synthesis models still have to
rely  on  simple  parametrisations   of  energy  or  angular  momentum
equations \citep{nelemansetal00-1, dewi+tauris00-1, nelemans+tout05-1,
politano+weiler06-1, politano+weiler07-1}.

A fundamental  problem in advancing our understanding  of CE evolution
and magnetic  braking in close  binaries is the shortage  of stringent
observational constraints that  can be used to test  and calibrate the
theory.  PCEBs  consisting of a white  dwarf and a  main sequence star
are  a very  well-suited  class  of objects  to  provide the  required
innovative   observational  input  since   they  are   numerous,  well
understood in terms  of their stellar components, they  are bright and
hence accessible  with 2--8 metre  telescopes, and their study  is not
complicated by mass transfer. \citet{schreiber+gaensicke03-1} analysed
a sample  of 30 well studied  PCEBs and showed that  the population of
PCEBs known so  far is not only small but  also heavily biased towards
young systems with low-mass secondary stars, and can hence not be used
for         comparison          with         population         models
\citep[e.g.][]{willems+kolb04-1}.

\begin{figure*}
\includegraphics[width=0.83\textwidth]{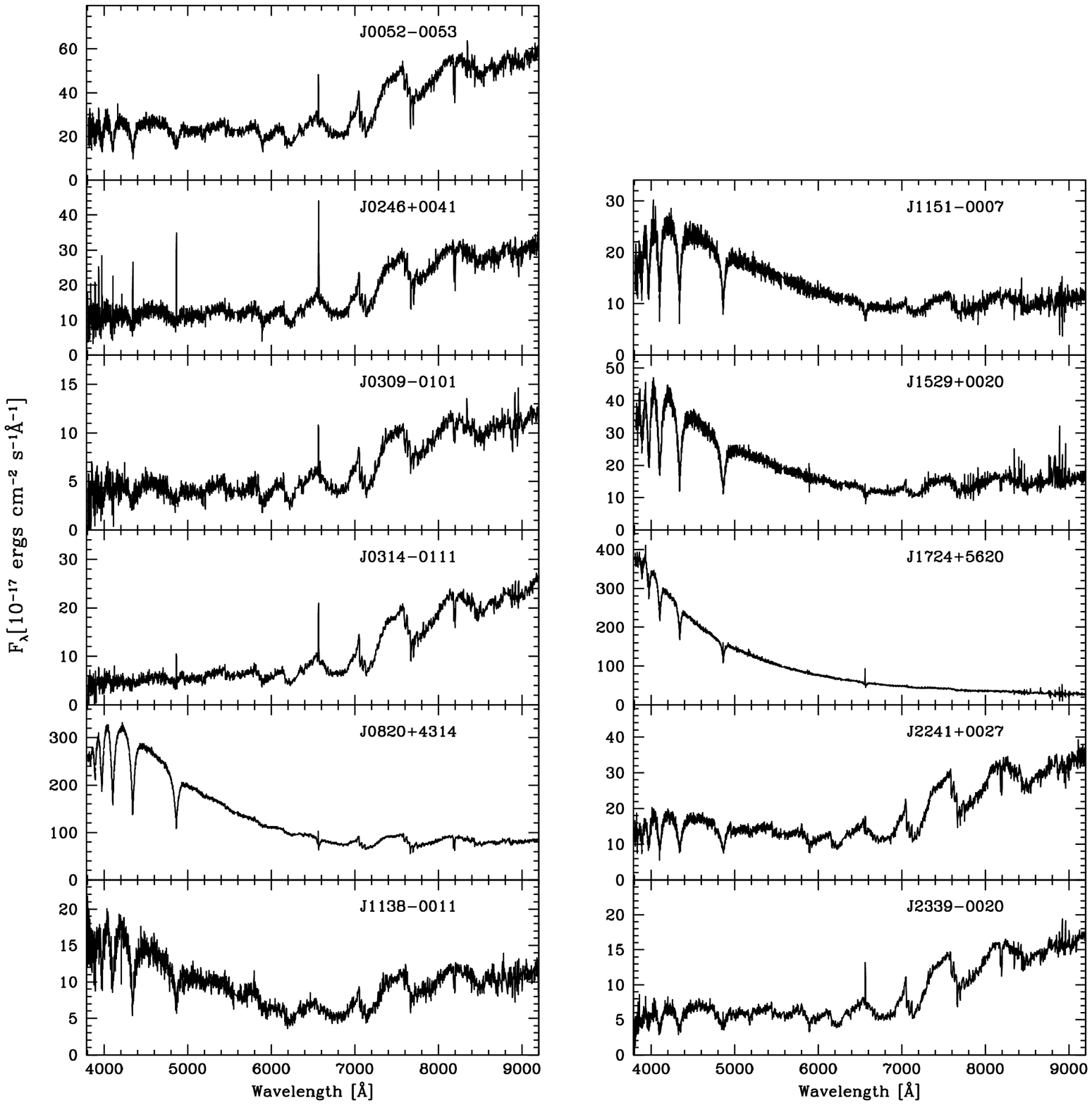} 
\caption{\label{f-spec} SDSS  spectra of the 11 SDSS  white dwarf main
sequence binaries studied in this work.}
\end{figure*}

The    Sloan    Digital    Sky    Survey    \citep[SDSS]{yorketal00-1,
stoughtonetal02-1, adelman-mccarthyetal08-1} is currently dramatically
increasing the number  of known white dwarf plus  main sequence (WDMS)
binaries \citep{silvestrietal07-1,  schreiberetal07-1}, paving the way
for  large-scale  observational  PCEB  population  studies.   We  have
initiated an  observational programme to identify the  PCEBs among the
SDSS  WDMS   binaries,  and  to  determine   their  binary  parameters
\citep{rebassa-mansergasetal07-1,  schreiberetal08-1}.   Those systems
in  which  significant  radial  velocity variation  is  detected,  are
classified  as close  binaries  or PCEBs.  Here  we present  follow-up
observations     of    11     PCEB     candidates    identified     in
\citet{rebassa-mansergasetal07-1} and  provide accurate values  of the
orbital periods as  well as estimates of their  stellar parameters and
orbital inclinations for seven of these systems.

\section{Observations}
\label{s-obs}

We have obtained time-resolved spectroscopy and photometry for 11 SDSS
WDMS binaries  (Table\,\ref{t-logobs}, Fig.\,\ref{f-spec}), henceforth
designated   SDSS\,J0052--0053,  SDSS\,J0246+0041,  SDSS\,J0309--0101,
SDSS\,J0820+4314,         SDSS\,J0314--0111,        SDSS\,J1138--0011,
SDSS\,J1151--0007,         SDSS\,J1529+0020,         SDSS\,J1724+5620,
SDSS\,J2241+0027 and SDSS\,J2339--0020.  Below we briefly describe the
instrumentation used  for the observations, and  outline the reduction
of the data.

\subsection{Spectroscopy}
\label{s-spectroscopy}

(i)    \textit{Magellan-Baade}.    Intermediate-resolution   long-slit
spectroscopy of SDSS\,J0314--0111 and SDSS\,J2241+0027 was obtained on
the  nights  of  2007  October   2  and  3  using  the  IMACS  imaging
spectrograph attached to the  Magellan Baade telescope at Las Campanas
Observatory. The 600\,$\ell$\,mm$^{-1}$ red-sensitive grating was used
along  with the  slit-view camera  and a  slit width  of 0.75\,arcsec,
giving  a reciprocal  dispersion of  0.39\,\AA\,px$^{-1}$.   The IMACS
detector is  a mosaic of  eight 2k$\times$4k SITe CCDs,  and long-slit
spectra with this instrument are spread over the short axis of four of
these  CCDs.  Using  a  grating  tilt of  14.7$^\circ$  allowed us  to
position the \Lines{Na}{I}{8183.27,8194.81} doublet towards the centre
of CCD2 and the H$\alpha$ line near the centre of CCD4.

\begin{table*}
  \caption[]{Radial       velocities      measured       from      the
    \Lines{Na}{I}{8183.27,8194.81}       doublet,      except      for
    SDSS\,J1724+5620,  where  radial   velocities  measured  from  the
    H$\alpha$  emission  line   in  the  15\,min  SDSS-subspectra  are
    given.  Note   that  the  provided   HJD  is  the   resulted  from
    HJD-2400000.}
\label{t-rvs}
\setlength{\tabcolsep}{0.8ex} 
\newcommand{\sd}[1]{\textit{#1}}
\begin{flushleft}
\begin{tabular}{lr@{\hspace*{4ex}}lr@{\hspace*{4ex}}lr@{\hspace*{4ex}}lr@{\hspace*{4ex}}lr}
\hline
\hline
\noalign{\smallskip}
HJD & RV [$\mathrm{km\,s^{-1}}$] & 
HJD & RV [$\mathrm{km\,s^{-1}}$] & 
HJD & RV [$\mathrm{km\,s^{-1}}$] & 
HJD & RV [$\mathrm{km\,s^{-1}}$] & 
HJD & RV [$\mathrm{km\,s^{-1}}$] \\
\noalign{\smallskip}
\hline
\noalign{\smallskip}
\multicolumn{2}{c}{\textbf{SDSS\,J0052--0053}} & 54379.8169 & 106.6 $\pm$   7.4                 & \multicolumn{2}{c}{\textbf{SDSS\,J1138--0011}} & 54238.7778 &  59.7 $\pm$  11.3                & 51993.9263 &  -81.3 $\pm$  10.7               \\
54328.7467 &  42.5 $\pm$  18.7                 & 54379.8622 &  42.5 $\pm$  14.5                 & 54205.5902 & -12.8 $\pm$   5.7                 & 54238.8109 &-156.4 $\pm$  11.2                & 51993.9387 & -101.9 $\pm$  10.4               \\
54328.7933 & -15.0 $\pm$  13.6                 & 54380.8165 & -97.1 $\pm$   9.9                 & 54207.7340 &  -7.7 $\pm$  12.6                 & 54238.8270 &-160.8 $\pm$  32.7                & 51993.9511 & -111.0 $\pm$  11.4               \\
54328.8077 & -40.7 $\pm$  13.7                 & 54380.8372 & -92.3 $\pm$   8.2                 & 54237.5647 & -13.9 $\pm$  11.6                 & 54238.8372 &-188.3 $\pm$  14.9                & 51994.9683 & -127.4 $\pm$  10.6               \\
54328.8825 &  31.5 $\pm$  27.5                 & 54380.8762 & -85.7 $\pm$   7.8                 & 54237.5721 &   5.1 $\pm$  10.3                 & 54239.5635 & 157.9 $\pm$   9.9                & 51994.9805 & -130.2 $\pm$  10.5               \\
54328.8969 &  -4.0 $\pm$   9.9                 & 54380.8842 & -72.9 $\pm$   8.0                 & 54238.4936 &   9.9 $\pm$   9.6                 & 54239.6756 &-121.3 $\pm$   8.0                & 51994.9928 & -131.6 $\pm$  12.5               \\
54329.6885 &  31.5 $\pm$   9.3                 & 54381.6121 & -63.7 $\pm$  12.7                 & 54238.5352 &  -5.1 $\pm$   8.5                 & 54240.5548 & 202.6 $\pm$  21.2                & 51997.9480 & -106.9 $\pm$  10.6               \\
54329.7028 & -12.1 $\pm$  11.8                 & 54381.7427 &  77.3 $\pm$   7.6                 & 54238.6889 &  23.1 $\pm$  10.6                 & 54240.6338 &-182.8 $\pm$   7.8                & 51997.9602 & -132.0 $\pm$  10.8               \\
54329.7172 & -50.9 $\pm$  16.4                 & 54381.8122 & 126.4 $\pm$   8.0                 & 54271.3996 &   9.2 $\pm$   6.7                 & 54240.6844 &  28.9 $\pm$  10.1                & 51997.9724 & -136.1 $\pm$  11.4               \\
54329.8269 & -56.1 $\pm$  14.1                 & 54381.8958 & 157.2 $\pm$   8.4                 & \multicolumn{2}{c}{\textbf{SDSS\,J1151--0007}} & 54240.7296 & 204.4 $\pm$   8.7                & \multicolumn{2}{c}{\textbf{SDSS\,J2241+0027}} \\
54329.8412 & -60.8 $\pm$  13.4                 & 54382.8362 & -31.9 $\pm$   8.9                 & 54237.5904 &  48.4 $\pm$  14.9                 & 54240.7357 & 194.9 $\pm$   7.9                & 54376.5842 &   2.6 $\pm$  14.0                \\
54329.8556 & -55.3 $\pm$  14.8                 & 54385.7884 & -71.1 $\pm$  11.5                 & 54237.5976 & 129.3 $\pm$  11.4                 & 54240.7417 & 175.1 $\pm$   9.5                & 54377.5637 &   6.6 $\pm$   6.6                \\
54330.7154 &  29.7 $\pm$  10.8                 & \multicolumn{2}{c}{\textbf{SDSS\,J0309--0101}} & 54238.5245 &-147.6 $\pm$  13.0                 & 54240.8311 &-114.7 $\pm$   8.1                & 54378.5290 &  13.2 $\pm$   9.9                \\
54330.7298 & -30.8 $\pm$  12.3                 & 54381.6726 &  60.4 $\pm$  12.4                 & 54238.5709 & -81.0 $\pm$  11.6                 & 54240.8372 & -87.6 $\pm$   9.9                & 54379.5523 &  39.9 $\pm$  13.3                \\
54330.7441 & -59.7 $\pm$  16.0                 & 54381.7714 &  31.5 $\pm$  12.0                 & 54238.7012 &-160.1 $\pm$   6.9                 & 54240.8432 & -54.6 $\pm$  12.7                & 54382.5261 & -20.1 $\pm$  17.7                \\
54330.8136 &  56.8 $\pm$  17.7                 & 54381.8386 &  41.0 $\pm$  10.7                 & 54239.5084 &-129.0 $\pm$   7.7                 & 54240.8498 &   8.4 $\pm$  31.7                & \multicolumn{2}{c}{\textbf{SDSS\,J2339--0020}} \\
54330.8279 &  28.2 $\pm$   9.3                 & 54382.8626 &  51.7 $\pm$  15.9                 & 54239.6047 & 233.4 $\pm$   8.0                 &            &                                  & 54380.5088 & -25.3 $\pm$  13.4                \\
54330.8423 & -29.3 $\pm$  12.2                 &            &                                   & 54239.6602 &-175.8 $\pm$   9.8                 & \multicolumn{2}{c}{\textbf{SDSS\,J1724+5620}} & 54380.5511 &   3.7 $\pm$   8.2                \\
54330.9158 &  32.6 $\pm$  14.1                 & \multicolumn{2}{c}{\textbf{SDSS\,J0314--0111}} & 54240.4793 & 148.4 $\pm$  11.7                 & 51812.6531 &  133.8 $\pm$  13.2               & 54381.5526 &-116.1 $\pm$   9.2                \\
54332.7313 &  22.7 $\pm$  11.1                 & 54376.7386 & -54.6 $\pm$   8.2                 & 54240.5236 &-227.1 $\pm$  12.5                 & 51812.6656 &  114.2 $\pm$  12.2               & 54382.5097 &  11.7 $\pm$  10.2                \\
54332.7457 &  59.3 $\pm$  23.5                 & 54376.7862 &-151.3 $\pm$   8.5                 & 54240.5291 &-207.4 $\pm$  11.5                 & 51813.5993 &   60.8 $\pm$  25.2               & 54382.5869 &  94.2 $\pm$   9.2                \\
54332.8433 &  23.1 $\pm$   9.6                 & 54376.8051 &-116.1 $\pm$  11.4                 & 54240.5346 &-198.2 $\pm$   8.5                 & 51813.6161 &   77.2 $\pm$  17.1               & 54382.6866 &  86.1 $\pm$  10.0                \\
           &                                   & 54376.8581 &  64.5 $\pm$  10.8                 & 54240.5400 &-163.8 $\pm$   9.5                 & 51813.6284 &  132.0 $\pm$  11.4               & 54382.7587 &  24.2 $\pm$   9.1                \\
\multicolumn{2}{c}{\textbf{SDSS\,J0246+0041}}  & 54376.8996 & 174.7 $\pm$  30.0                 & 54240.6033 & 219.8 $\pm$  11.4                 & 51813.6425 &  118.3 $\pm$  12.3               & 54383.5274 &-134.4 $\pm$  11.2                \\
54378.6386 &-127.5 $\pm$  11.4                 & 54377.6572 & 119.8 $\pm$   8.0                 & 54240.6088 & 220.5 $\pm$   9.3                 & 51813.6631 &  121.1 $\pm$  11.1               & 54385.5605 &-181.3 $\pm$  17.0                \\
54378.6496 &-122.7 $\pm$  13.0                 & 54377.6830 & 186.1 $\pm$   9.3                 & 54240.6142 & 161.2 $\pm$  10.1                 & 51813.6768 &  103.2 $\pm$  10.6               & 54385.6462 &-177.3 $\pm$   9.0                \\
54378.7955 &  18.3 $\pm$   8.8                 & 54377.7380 & 164.5 $\pm$   8.0                 & 54240.6197 & 111.7 $\pm$   9.3                 & 51813.6889 &   94.1 $\pm$  10.7               & 54385.6663 &-164.9 $\pm$  23.9                \\
54379.6704 & 176.9 $\pm$  11.3                 & 54377.7782 &  -2.6 $\pm$   7.5                 &            &                                   & 51818.5877 &   71.7 $\pm$  22.8               & 54385.6864 &-133.4 $\pm$  12.8                \\
54379.6789 & 159.0 $\pm$  15.6                 & 54377.8165 &-116.1 $\pm$   9.5                 & \multicolumn{2}{c}{\textbf{SDSS\,J1529+0020}}  & 51818.6178 &  109.6 $\pm$  13.0               & 54385.7065 &-104.0 $\pm$  21.6                \\
54379.6875 & 177.3 $\pm$   9.7                 & 54377.8672 & -98.9 $\pm$   8.1                 & 54238.6219 &  -3.3 $\pm$  23.9                 & 51818.6297 &  122.4 $\pm$  13.9               & 54385.7495 & -49.8 $\pm$  27.1                \\
54379.7813 & 132.6 $\pm$   7.9                 & 54377.8866 & -44.3 $\pm$  12.1                 & 54238.6795 &-145.1 $\pm$  11.7                 & 51818.6418 &  127.4 $\pm$  13.2               & 54385.7696 & -16.9 $\pm$   9.8                \\
\hline\noalign{\smallskip}
\end{tabular}
\end{flushleft}
\end{table*}

The images  were reduced using  {\sc starlink} software, and  the {\sc
pamela} and  {\sc molly}  packages \citep{marsh89-1} were  employed to
optimally  extract and  calibrate  the spectra.   By fitting  Gaussian
functions to arc and sky emission  lines we find that the spectra have
a  resolution of  approximately 1.6\,\AA.   Helium arc  lamp exposures
were taken at the start of  each night in order to derive a wavelength
solution   with  a   statistical  uncertainty   of   only  0.002\,\AA\
\citep{marshetal94-1}. This was applied  to each spectrum taken on the
same  night.   Detector flexure  was  measured  and  removed from  the
wavelength  solution for  each  spectrum using  the  positions of  sky
emission  lines  \citep{southworthetal06-1, schreiberetal08-1}.   Flux
calibration and telluric line  removal was performed using {\sc molly}
and spectra of the standard star BD\,+28$^\circ$4211.

(ii)  \textit{Magellan-Clay  and  Very  Large Telescope  (VLT)}.   The
observations     of     SDSS\,J1138-0011,    SDSS\,J1151--0007     and
SDSS\,J1529+0020 were obtained with the  same setup and reduced in the
same manner  as described  by \citet{schreiberetal08-1}, and  we refer
the reader to that paper for full details.

(iii) \textit{New Technology Telescope (NTT)}. Two observing runs were
carried  out  at  the  NTT  in  August  and  October  2007,  providing
intermediate   time-resolved    spectroscopy   of   SDSS\,J0052--0053,
SDSS\,J0246+0041,      SDSS\,J0309-0101,      SDSS\,J2241+0027     and
SDSS\,J2339--0020.  We  used the  EMMI spectrograph equipped  with the
Grat\#7 grating, the MIT/LL red  mosaic detector, and a 1\,arcsec wide
long-slit,     resulting    in     a     wavelength    coverage     of
$\lambda7770-8830$\,\AA.     The   data   were    bias-corrected   and
flat-fielded  using  the {\sc  starlink}  software,  and spectra  were
optimally extracted  using the {\sc pamela}  package.  Helium-Agon arc
lamp spectra were  taken at the beginning of each  night and were then
used  to  establish a  generic  pixel-wavelength  relation using  {\sc
molly}. Specifically,  we fitted a fourth order  polynomial which gave
rms smaller than 0.02\AA\ for all spectra.  We then used the night sky
emission lines to adjust the zero-point of the wavelength calibration,
thus correcting  for instrument  flexure.  The spectral  resolution of
our  instrumental  setup determined  from  the  sky  lines is  2.8\AA.
Finally,  the  spectra  were  calibrated and  corrected  for  telluric
absorption within {\sc molly}  using observations of the standard star
Feige\,110.

(iv)  \textit{William  Herschel  Telescope  (WHT).}  One  spectrum  of
SDSS\,J1138-0011 was taken with  the 4.2\,m William Herschel Telescope
in June  2007 at the Roque  de los Muchachos observatory  on La Palma.
The double-beam ISIS spectrograph was  equipped with the R158R and the
R300B  gratings, and  a  1\,arcsec long-slit,  providing a  wavelength
coverage of $\lambda7600-9000$\,\AA.  The spectral resolution measured
from the sky lines is 1.6\,\AA. Reduction and calibration were carried
in the same way as described for the NTT above.

(v) \textit{SDSS.}  For one  system, SDSS\,J1724+5620, we were able to
determine an  accurate orbital period  from our photometry  alone, but
were  lacking follow-up  spectroscopy.  SDSS\,DR6  contains  three 1-d
calibrated  spectra for  SDSS\,J1724+5620  (MJD-PLT-FIB 51813-357-579,
51818-358-318, and  51997-367-564). Each of these  spectra is combined
from  at  least 3  individual  exposures  of  900\,s, which  are  also
individually           released          as           part          of
DR6\footnote{http://www.sdss.org/dr6/dm/flatFiles/spCFrame.html}.  For
SDSS\,J1724+5620, a total of 23 sub-spectra are available, of which 22
allowed reliable radial velocity measurements.

\begin{figure*}
\includegraphics[width=0.47\textwidth]{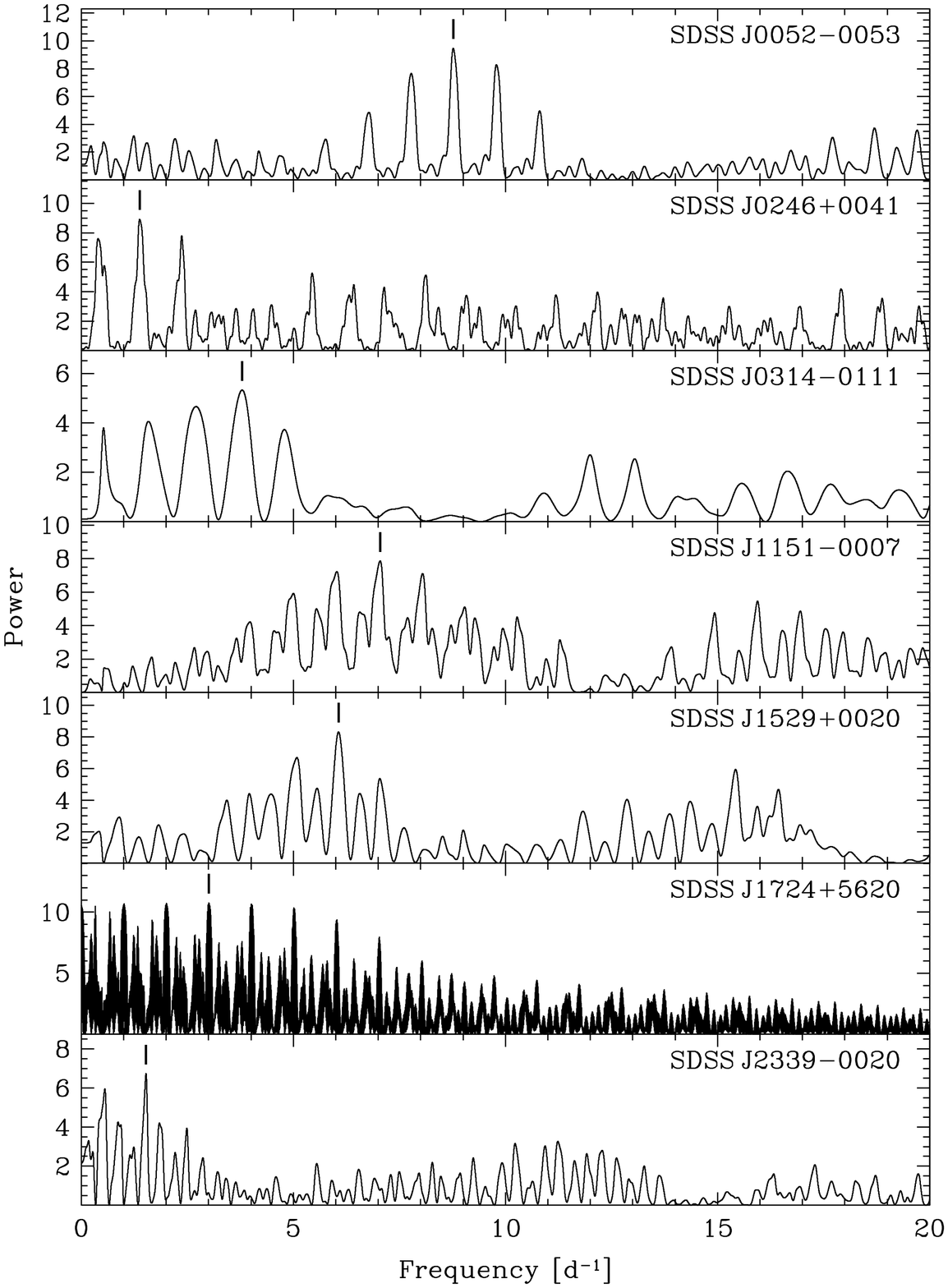} 
\hspace*{4ex}
\includegraphics[width=0.47\textwidth]{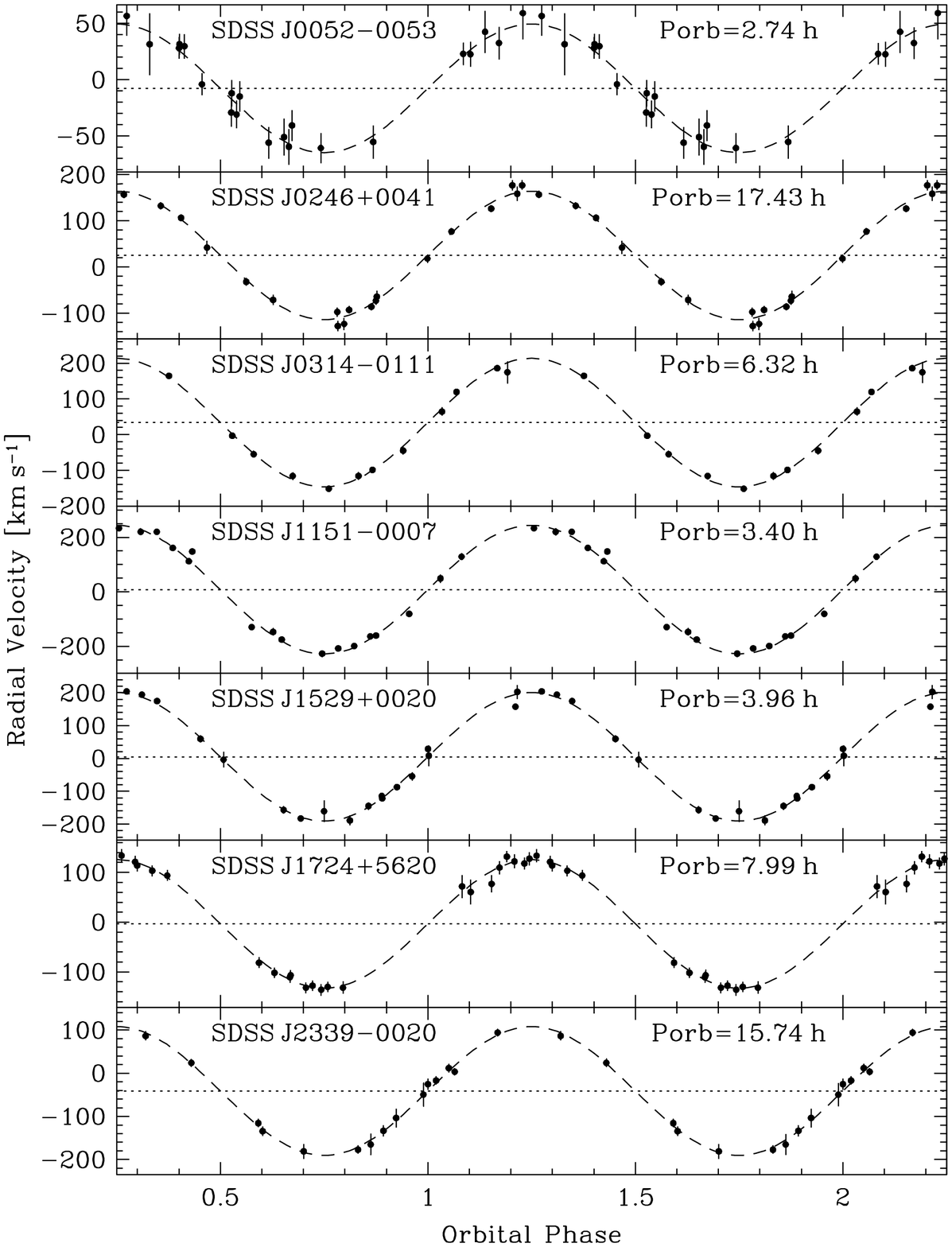}
\caption{\label{f-scargle} Left:  Scargle periodograms calculated from
  the  radial  velocity   variations  measured  from  the  \Ion{Na}{I}
  absorption    doublet   in    SDSS\,J0052--0053,   SDSS\,J0246+0041,
  SDSS\,J0314--0111,    SDSS\,J1151--0007,    SDSS\,J1529+0020,    and
  SDSS\,J2339--0020,   and  from  the   H$\alpha$  emission   line  in
  SDSS\,J1724+5620. The aliases with the highest power are indicated
  by tick marks.  Right: the  radial velocity  measurements folded
  over the orbital periods of the systems, as determined from the best
  sine fits  (Table\,\ref{t-periodaliases}).}
\end{figure*}

\subsection{Photometry}

(i)   \textit{IAC80  and   AIP\,70\,cm   telescopes.}   We   obtained
differential photometry  of SDSS\,J1724+5620 with  the IAC80 telescope
at the Observatorio del Teide  (Spain) and the 70\,cm telescope of the
Astrophysical Institute of Postdam at Babelsberg (Germany) for a total
of 13 nights during May, August and September 2006 and March 2007. The
IAC80 cm  telescope was  equipped with a  2k$\times$2k CCD.  A binning
factor of 2  was applied in both spatial directions  and only a region
of 270$\times$270  (binned) pixels of  size 0.66 arcsec was  read. The
detector  used at  the  Babelsberg  70 cm  telescope  was a  cryogenic
1k$\times$1k Tek CCD.  The whole frame was read  with a binning factor
of  3, resulting  in a  scale  of 1.41  arcsec/pixel. A  semiautomated
pipeline involving  DoPHOT was used  to reduce the images  and extract
the photometric information.

(ii) \textit{Calar Alto 2.2\,m telescope.} We used CAFOS with the SITe
2k$\times$2k pixel CCD camera on  the 2.2 metre telescope at the Calar
Alto  observatory  to obtain  filter-less  differential photometry  of
SDSS\,J0314--0111. Only a small part of  the CCD was read out in order
to  improve the  time resolution.   The  data were  reduced using  the
pipeline  described in \citet{gaensickeetal04-1},  which pre-processes
the raw  images in  MIDAS and extracts  aperture photometry  using the
\textsc{Sextractor} \citep{bertin+arnouts96-1}.

(iii)  \textit{Kryoneri 1.2\,m  telescope}  Filter-less photometry  of
SDSS\,J0820+4311 was obtained in  November 2006 at the 1.2\,m Kryoneri
telescope using a Photometrics SI-502 $516\times516$ pixel camera. The
data reduction was carried out in  the same way as described above for
the Calar Alto observations.

\section{Orbital Periods}
\label{s-periods}

\subsection{Radial Velocities}
\label{s-rvs}

In \citet{rebassa-mansergasetal07-1} we measured the radial velocities
fitting a second order  polynomial plus a double-Gaussian line profile
of fixed  separation to the  \Lines{Na}{I}{8183.27,8194.81} absorption
doublet.  Free  parameters were  the amplitude and  the width  of each
Gaussian and  the radial  velocity of the  combined doublet.   Here we
adopt  a  slightly  modified  approach,  using  just  a  single  width
parameter for both  line components.  This reduction in  the number of
free  parameters  increases  the   robustness  of  the  fits.   Radial
velocities measured  in this way for  nine WDMS binaries  are given in
Table\,\ref{t-rvs}. In addition, we measured the companion star radial
velocities  for SDSS\,J1724+5620  by means  of a  Gaussian fit  to the
H$\alpha$   emission  line  clearly   visible  in   22  of   the  SDSS
sub-exposures  (Sect.\,\ref{s-spectroscopy}), which are  also reported
in Table\,\ref{t-rvs}.

\citet{scargle82-1} periodograms calculated from the radial velocities
of  each system  to investigate  the periodic  nature of  the velocity
variations contain a number of  aliases due to the sampling pattern of
the observations (Fig.\,\ref{f-scargle},  left panel).  We carried out
sine-fits of the form

\begin{equation}
\label{e-fit}
V_\mathrm{r} =
K_\mathrm{sec}\,\sin\left[\frac{2\pi(t-T_0)}{P_\mathrm{orb}}\right]
+\gamma
\end{equation}

\noindent
to  the velocity  data  sets of  each  system, where  $\gamma$ is  the
systemic   velocity,   $K_\mathrm{sec}$   is   the   radial   velocity
semi-amplitude of  the companion star,  $T_0$ is the time  of inferior
conjunction of the secondary star, and $P_\mathrm{orb}$ is the orbital
period.    Several   fits   were   done,  adopting   the   frequencies
corresponding to the  strongest peaks in the power  spectra as initial
conditions.  The parameters resulting  from these fits are reported in
Table\,~\ref{t-periodaliases}.          For         SDSS\,J0052--0053,
SDSS\,J0246+0041,  SDSS\,J0314--0111, SDSS1151--0007, SDSS\,J1529+0020
and  SDSS\,J2339--0020 the  sine-fits  allow a  unique  choice of  the
orbital  period.   For  SDSS\,J1724+5620,  the sampling  of  the  SDSS
spectra  is  very  sparse,  resulting  in  finely  structured  aliases
superimposed on a  sequence of large aliases spaced  by 1\,\id. A sine
fit to the radial velocities started off at the 3\,\id\ alias provides
a spectroscopic orbital period  of 7.99243(16)\,h, which is consistent
with  the   more  accurate   value  determined  from   the  photometry
(Sect.\,~\ref{s-light_curves}). The radial velocity data for all seven
systems   folded   over   their   orbital   periods   are   shown   in
Fig.\,~\ref{f-scargle} (right panel).

The periodograms  calculated from the radial  velocity measurements of
the   PCEB   candidates   SDSS\,J0309--0101,  SDSS\,J1138--0011,   and
SDSS\,J2241+0027 do not reveal any significant peak. The low amplitude
of  the radial  velocity variations  observed in  these  three objects
suggests that they  may be wide WDMS binaries  rather than PCEBs. This
issue will be discussed further in Sect.\,\ref{s-nonPCEB}.

\begin{table}
\setlength{\tabcolsep}{1ex}
\caption{\label{t-periodaliases}   Orbital   periods,  semi-amplitudes
  $K_\mathrm{sec}$,  systemic  velocities  $\gamma_\mathrm{sec}$,  and
  reduced $\chi^2$  from sine fits  to the \Ion{Na}{I}  doublet radial
  velocity  data  for  the  strongest  two to  three  aliases  in  the
  periodograms shown in Fig.\,\ref{f-scargle}. The best-fit values are
  set   in   bold.    The   SDSS  H$\alpha$   radial   velocities   of
  SDSS\,J1724+5620  are  folded  on  the  photometric  orbital  period
  obtained in Sect.\,\ref{s-light_curves} (which is more accurate than
  the spectroscopically determined value  of the orbital period). Note
  also  that   the  semi-amplitud   measurement  of  this   system  is
  underestimated,  as  it  comes  from RV  measurements  of  H$\alpha$
  emission   from  the   irradiated   face  of   the  companion   (see
  Sect.\,\ref{s-sdss1724}).}
\begin{flushleft}
\begin{center}
\begin{tabular}{lrrrr}\hline\hline
System & 
\multicolumn{1}{c}{\Porb} & 
\multicolumn{1}{c}{$K_\mathrm{sec}$} & 
\multicolumn{1}{c}{$\gamma_\mathrm{sec}$} & 
$\chi^2$  \\
& \multicolumn{1}{c}{[h]} & 
\multicolumn{1}{c}{$\mathrm{[km\,s^{-1}]}$} & 
\multicolumn{1}{c}{$\mathrm{[km\,s^{-1}]}$} & \\
\hline
\noalign{\smallskip}
\textbf{SDSS\,J0052--0053} & 
  3.0850$\pm$0.0090 & 47.2$\pm$6.6  & -2.9$\pm$5.8 & 3.74 \\
& \textbf{2.7350$\pm$0.0023} & \textbf{57.0$\pm$3.1} & \textbf{-7.2$\pm$2.6} & \textbf{0.50} \\
& 2.4513$\pm$0.0035 & 54.8$\pm$5.8 & -9.1$\pm$4.6 & 1.96 \\
\textbf{SDSS\,J0246+0041}    & 
61.1$\pm$2.7     & 124$\pm17$    & 16$\pm$12    & 20.3 \\
& \textbf{17.432$\pm$0.036} & \textbf{140.7$\pm$3.5} & \textbf{24.9$\pm$2.7} & \textbf{1.25} \\
& 10.130$\pm$0.031 & 163$\pm$16    & 34$\pm$10    & 21.6 \\
\textbf{SDSS\,J0314--0111}   & 
8.66$\pm$0.17   & 154$\pm$19    & 44$\pm$16   & 27.1 \\
& \textbf{6.319$\pm$0.015} & \textbf{174.9$\pm$4.8} & \textbf{31.0$\pm$3.4} & \textbf{1.15} \\
\textbf{SDSS\,J1151--0007}   & 
3.979$\pm$0.013   & 202$\pm$20    & -2$\pm$16    & 34.4 \\
& \textbf{3.3987$\pm$0.0027} & \textbf{233.8$\pm$8.1} &  \textbf{9.0$\pm$5.9} & \textbf{3.55} \\
& 2.9849$\pm$0.0092 & 213$\pm$22    & -22$\pm$23   & 64.7 \\
\textbf{SDSS\,J1529+0020}    & 
4.759$\pm$0.028   & 171$\pm$18    & 11$\pm$14   & 26.9 \\
&\textbf{3.9624$\pm$0.0033} & \textbf{193.1$\pm$5.2} & \textbf{6.0$\pm$4.1} & \textbf{1.91} \\
&1.5558$\pm$0.0024 & 182$\pm$21    & 6$\pm$15    & 36.7 \\
\textbf{SDSS\,J1724+5620}    &  \textbf{7.992463(31)} & \textbf{129.3$\pm$1.9} & \textbf{-3.5$\pm$1.8} & \textbf{0.41} \\
\textbf{SDSS\,J2339--0020}   & 
43.18$\pm$0.78  & 121$\pm$18  & -50$\pm$15 & 13.5 \\
& \textbf{15.744$\pm$0.016} & \textbf{149.8$\pm$4.0} & \textbf{-40.9$\pm$2.6}  & \textbf{0.69} \\
\hline\noalign{\smallskip}
\end{tabular}
\end{center}
\end{flushleft}
\end{table}

\subsection{Light curves}
\label{s-light_curves}

The  light curves  of SDSS\,J0314--0111  and  SDSS\,J1724+5620 display
variability with  an amplitude of  $\sim0.05$\,mag and $\sim0.1$\,mag,
respectively.  We  calculated periodograms for both  systems using the
\textsf{ORT/TSA} command in \textsf{MIDAS}, which folds and phase-bins
the data  using a grid of trial  periods and fits a  series of Fourier
terms to the folded light curve \citep{schwarzenberg-czerny96-1}.

\begin{figure*}
\includegraphics[width=0.63\textwidth]{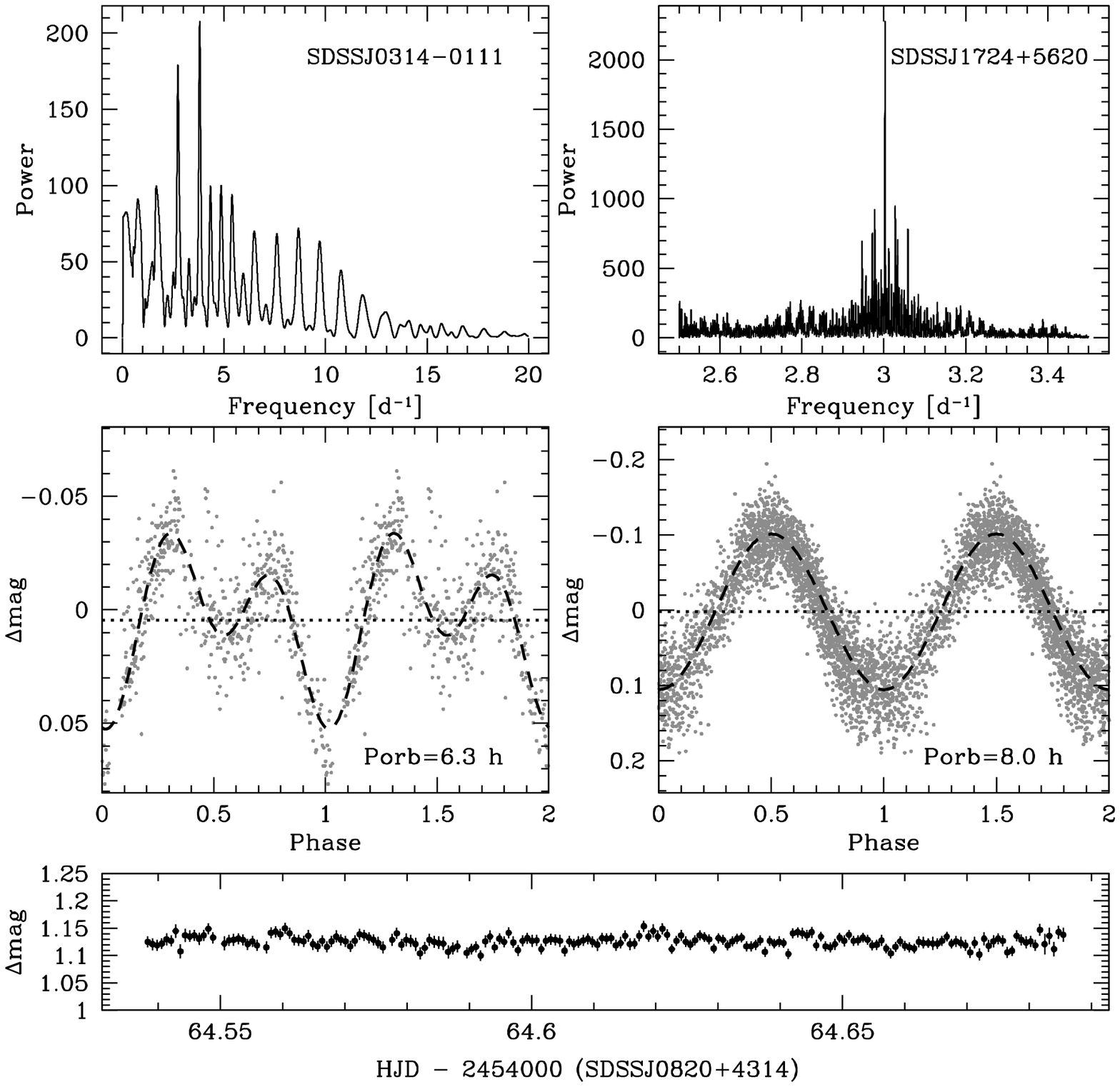} 
\caption{\label{f-phot}   Top    and   middle   panels:   \textsf{ORT}
  periodograms  and  light  curves  folded  on  strongest  photometric
  periodicities,  $\Porb=6.32\pm0.02$\,h   for  SDSS\,J0314--0111  and
  $\Porb=7.99246\pm0.00003$\,h  for  SDSS\,J1724+5620.  Bottom  panel:
  the  light curve  of  SDSS\,J0820+4314 is  essentially  flat, and  a
  periodogram  calculated   from  this  data  does   not  contain  any
  significant periodicity.}
\end{figure*}

A  strong peak  is found  in the  periodogram of  SDSS\,J0314--0111 at
3.8\,\id,  i.e. 6.3\,h (Fig.\,\ref{f-phot},  top left  panel). Folding
the  photometric data  over  that period  results  in a  double-humped
modulation    which   we    identify    as   ellipsoidal    modulation
(Fig.\ref{f-phot},  middle left panel).  The detection  of ellipsoidal
modulation  indicates  that  the  companion  star must  be  filling  a
significant amount of its Roche-lobe radius and have a moderately high
inclination,  both   of  which  hypothesis  are   confirmed  below  in
Sect.\,~\ref{s-param}.  The  two minima  differ in depth,  as expected
for ellipsoidal  modulation because of the  stronger gravity darkening
on  the hemisphere facing  the white  dwarf. The  two maxima  are also
unequal, which is  observed relatively often in PCEBs,  and thought to
be     related      to     the     presence      of     star     spots
\citep[e.g.][]{kawka+vennes03-1,tappertetal07-1}. We conclude that the
strongest periodicity detected  in the photometry of SDSS\,J0314--0111
is    consistent    with     the    spectroscopic    orbital    period
(Table\,\ref{t-periodaliases}). As the radial velocities span a longer
temporal baseline than the  photometry, they provide the more accurate
period   measurement,  and   we  adopt   $\Porb=6.319\pm0.015$\,h  for
SDSS\,J0314--0111.

\begin{figure}
\includegraphics[width=\columnwidth]{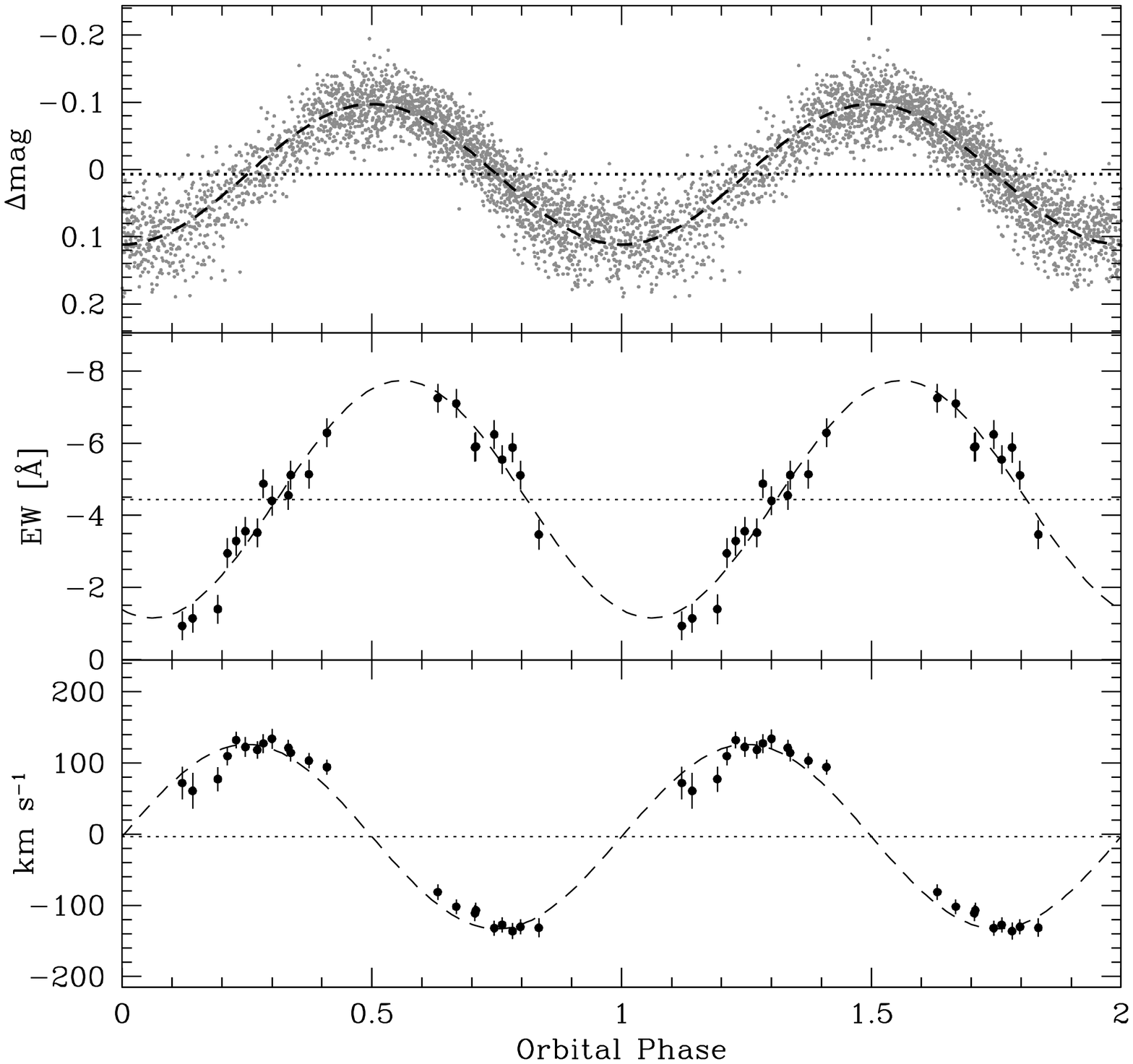}
\caption{\label{f-rvphot1724} Top  panel: The IAC\,80  and AIP\,70\,cm
  photometry of SDSS\,J1724+5620 folded over the photometric period of
  $7.9924632\pm0.0000312$\,h.   Middle   pane:  the  equivalent  width
  variation  of the  H$\alpha$ emission  line measured  from  the SDSS
  spectra  of SDSS\,J1724+5620  (Table\,~\ref{t-rvs}) folded  over the
  photometric period. Maximum equivalent  width occur roughly in phase
  with the maximum in the light curve (see Sect.\,\ref{s-sdss1724} for
  details).   Bottom  panel:  the  radial velocity  variation  of  the
  H$\alpha$ emission  line. The relative  phasing with respect  to the
  photometry  is  consistent  with  the photometric  modulation  being
  caused by irradiation of the secondary star by the hot white dwarf.}
\end{figure}

For the analysis of SDSS\,J1724+5620 we separately merged the data for
the $R$  and $I$ band, normalised  both data sets  to their respective
mean values, and finally combined all data into a single data set. The
\textsf{ORT} periodogram calculated from the full light curve displays
a strong signal  at 3\,\id, with weak aliases  related to the sampling
pattern inherent  to the  data (Fig.\,\ref{f-phot}, top  right panel).
Given the large quasi-sinusoidal shape of the modulation, and the high
effective temperature  of the white  dwarf (Table\,~\ref{t-param}), we
identify the observed photometric  variability as being due to heating
effect  (or  reflection effect  as  it is  often  referred  to in  the
literature).  A sine  fit to the combined data  (dashed line) provided
the following photometric ephemeris

\begin{equation}
\mathrm{HJD} (\phi=0) = 2453865.29(1) + \mathrm{E}\times0.3330193(13),
\label{e-ephem}
\end{equation}

\noindent
where  $\phi =  0$ refers  to the  occurrence of  minimum  light.  The
photometric period of 8\,h is consistent with the alias pattern in the
periodogram calculated  from the H$\alpha$  radial velocity variations
(Fig.\,~\ref{f-scargle},  left panel).   The phase-folded  light curve
using the photometric  period is shown on the  middle right hand panel
of Fig.\,\ref{f-phot}.  Based on  the above ephemeris, the accumulated
phase  error for  the  first SDSS  subspectrum (Table~\ref{t-rvs})  is
0.024  cycles,  which  is   sufficiently  good  to  phase  the  radial
velocities     obtained     from     the     SDSS     spectra     with
Eq.\,(\ref{e-ephem}). The phase-folded light curve and radial velocity
curve are shown in  Fig.\,\ref{f-rvphot1724}. The phase offset between
the  two   curves  is   $0.226\pm0.005$,  which  is   consistent  with
$\simeq0.25$ within  the error of the  fit and the  ephemeris.  A 0.25
phase  offset  is  what  is  expected from  the  assumption  that  the
photometric modulation  is related to a heating  effect, i.e.  maximum
light and  red-to-blue crossing of the radial  velocity corresponds to
the superior conjunction of  the secondary star, whereas minimum light
and  blue-to-red crossing of  the radial  velocity corresponds  to its
inferior conjunction.

Finally, we show in the bottom panel of Fig.\,\ref{f-phot} a 3.5 hours
light  curve of SDSS\,J0820+4314.  We monitored  the system  through a
total    of   8.2    hours    on   three    different   nights    (see
Table\,\ref{t-logobs}),   and   found   no   significant   photometric
modulation.   This implies  that the  companion star  is significantly
under-filling  its Roche lobe  and/or that  the binary  inclination is
very low.   The orbital  period of  the system will  hence need  to be
measured from a radial velocity study.

\section{Binary parameters}
\label{s-param}

In   \citet{rebassa-mansergasetal07-1}   we   developed   a   spectral
decomposing/fitting technique  for the analysis of  WDMS binaries with
SDSS spectroscopy. In brief,  this analysis determines the white dwarf
effective temperature, surface gravity,  mass and radius, the spectral
type  and  radius of  the  main sequence  companion,  as  well as  two
independent  distance estimates  based on  the properties  of  the two
stellar components. Given that the SDSS spectra reduction pipeline has
been  improved  with  DR6  \citep{adelman-mccarthyetal08-1},  we  have
re-analysed here the  seven PCEBs for which we  were able to determine
orbital  periods. Comparison  with the  results from  DR5  reported in
\citet{rebassa-mansergasetal07-1} shows that the fit parameters differ
slightly,  but  agree  in  the  vast  majority  of  cases  within  the
errors. We report the  average stellar parameters determined from fits
to  the  multiple  SDSS  spectra in  Table\,\ref{t-param}.   Rewriting
Kepler's third law (assuming common notation),

\begin{equation}
\frac{(M_\mathrm{wd}\sin i)^3}{(M_\mathrm{wd}+M_\mathrm{sec})^2}
=\frac{\Porb K_\mathrm{sec}^3}{2\pi G}
\end{equation}
\noindent as
\begin{equation}
\sin i = \frac{K_\mathrm{sec}}{M_\mathrm{wd}} \left(\frac{\Porb}
{2\pi G}\right)^{1/3}(M_\mathrm{wd}+M_\mathrm{sec})^{2/3},
\label{e-incl}
\end{equation}
\noindent
and adopting  the orbital period  determined from the analysis  of the
radial velocities and  the photometry, as well as  the masses from the
analysis of the SDSS spectra, we  are now able to estimate the orbital
inclinations  for  the  seven  systems in  Table\,\ref{t-param}.   Two
systems require some additional notes: SDSS\,J0314--0111 contains a DC
white dwarf, and  hence no white dwarf parameters  could be determined
from the spectral  analysis, and we assume $M_\mathrm{wd}=0.65M_\odot$
for the estimate of the inclination. In SDSS\,J1724+5620 the hot white
dwarf  is irradiating  the companion  star, and  the  implications are
discussed in more detail in Sect.\,\ref{s-sdss1724}.

Inspecting   Eq.(\ref{e-incl}),  it   is  clear   that   the  dominant
uncertainties  in the inclination  estimates are  only $M_\mathrm{wd}$
and  $M_\mathrm{sec}$,  as the  orbital  periods and  $K_\mathrm{sec}$
velocities are accurately determined.   The primary uncertainty in the
inclination  estimates is  $M_\mathrm{sec}$,  as it  is  based on  the
spectral   type  of   the  companion   star,  adopting   the  spectral
type-mass-radius relation  given by \citet{rebassa-mansergasetal07-1}.
$M_\mathrm{wd}$ is  determined from fitting the  Balmer line profiles,
and   relatively  well   constrained.   Given   that   $\sin  i\propto
M_\mathrm{sec}^{2/3}$, but $\sin i\propto M_\mathrm{wd}$, the relative
weight  of  the uncertainty  in  $M_\mathrm{sec}$  is alleviated.   We
estimate   the   uncertainties   on   the   binary   inclinations   in
Table\,\ref{t-param}  by assuming  in Eq.(\ref{e-incl})  the  range in
$M_\mathrm{sec}$ implied  by a  spectral type uncertainty  of $\pm0.5$
spectral classes plus the  associated errors in the spectral type-mass
relation, as well  as the range in $M_\mathrm{wd}$  resulting from the
Balmer  line profile  fits.  Given  the high  estimate for  the binary
inclination,  SDSS\,J1529+0020  is a  good  candidate for  photometric
follow-up observations probing for eclipses in its light curve.

\begin{table*}
\caption{\label{t-param}  Average binary  parameters obtained  for the
  seven   PCEBs  which  have   orbital  period   and  $K_\mathrm{sec}$
  measurements.  $M_\mathrm{wd}$,  $M_\mathrm{sec}$, $R_\mathrm{sec}$,
  spectral  type,   \Teff\,  and  $\log  g$   are  obtained  following
  \citet{rebassa-mansergasetal07-1},   except  for  SDSS\,J0314--0111,
  where we assume a WD mass  of 0.65 $\pm$ 0.1 $M_\odot$ (see text for
  details).    \Porb\,   and    $K_\mathrm{sec}$   are   measured   in
  Sect.\,\ref{s-periods}   of   this   work.  Estimates   of   orbital
  separations $a$,  $q$, $K_\mathrm{wd}$, secondary  Roche lobe radius
  R$_\mathrm{L_\mathrm{sec}}$ and  inclinations are obtained  from the
  equations  given  in  Sect.\,\ref{s-param}.  SDSS\,J1724+5620  is  a
  particular  case,  as  the  inner  hemisphere of  the  companion  is
  irradiated.   Constraints  on its  orbital  parameters are  obtained
  assuming  a spectral  type between  M3-5, and  considering different
  K$_\mathrm{sec,corr}$   values  for  each   mass  and   radius  (see
  Sect.\,\ref{s-sdss1724}).   An   additional   constraint   for   its
  inclination  comes  from  the  fact  that  SDSS\,J1724+5620  is  not
  eclipsing.}
\begin{flushleft}
\begin{center}
\setlength{\tabcolsep}{0.7ex}
\begin{tabular}{lrclrclrclrclrclrclrcl}\hline\hline
\noalign{\smallskip}
 & \multicolumn{3}{c}{SDSS\,J0052--0053} 
 & \multicolumn{3}{c}{SDSS\,J0246+0041}
 & \multicolumn{3}{c}{SDSS\,J0314--0111}
 & \multicolumn{3}{c}{SDSS\,J1151--0007}
 & \multicolumn{3}{c}{SDSS\,J1529+0020}
 & \multicolumn{3}{c}{SDSS\,J1724+5620}
 & \multicolumn{3}{c}{SDSS\,J2339--0020} \\
\hline
$M_\mathrm{wd} [M_\odot]$                  &    1.2&$\pm$&0.4    &   0.9&$\pm$&0.2   &  0.65&$\pm$& 0.1  &    0.6&$\pm$&0.1    &  0.40&$\pm$&0.04  &      0.42&$\pm$&0.01              &  0.8&$\pm$&0.4     \\
$M_\mathrm{sec} [M_\odot]$                 &    0.32&$\pm$&0.09  &   0.38&$\pm$&0.07  &  0.32&$\pm$&0.09 &    0.19&$\pm$&0.08  &  0.25&$\pm$&0.12  & $\sim$0.25&  -  &0.38             &  0.32&$\pm$&0.09   \\
$q$                                        &    0.3&$\pm$&0.1    &   0.4&$\pm$&0.1   &   0.5&$\pm$&0.2   &    0.3&$\pm$&0.2    &   0.6&$\pm$&0.3   & $\sim$0.6&  -  &0.9               &  0.4&$\pm$&0.2     \\
$a [R_\odot]$                              &    1.1&$\pm$&0.1    &   3.7&$\pm$&0.2   &   1.7&$\pm$&0.1   &    1.0&$\pm$&0.1    &   1.1&$\pm$&0.1   & $\sim$1.8&  -  &1.9               &  3.3&$\pm$&0.4     \\
\Porb\ [h]                                 &  2.735&$\pm$&0.002  & 17.43&$\pm$&0.04  &  6.32&$\pm$&0.02  &  3.399&$\pm$&0.003  & 3.962&$\pm$&0.003 &\multicolumn{3}{c}{7.992463(3)}    &15.74&$\pm$&0.02    \\
$K_\mathrm{sec} [\kms]$                    &     57&$\pm$& 3     &   141&$\pm$& 4    &   175&$\pm$& 5    &    234&$\pm$& 8     &   193&$\pm$& 5    & $\sim$129&  -  &214               &  150&$\pm$&4       \\
$K_\mathrm{wd} [\kms]$                     &     15&$\pm$& 6     &    59&$\pm$& 15   &    86&$\pm$&28    &     79&$\pm$&38     &   123&$\pm$&61    &  $\sim$78&  -  &194               &   57&$\pm$&29      \\
Sp$_\mathrm{sec}$                          &      4&$\pm$& 0.5   &     3&$\pm$& 0.5  &     4&$\pm$& 0.5  &      6&$\pm$& 0.5   &     5&$\pm$&0.5   &         3&  -  &5                 &     4&$\pm$&0.5    \\
$R_\mathrm{sec} [R_\odot]$                 &   0.33&$\pm$&0.10   &  0.39&$\pm$&0.08  &  0.33&$\pm$&0.10  &   0.19&$\pm$&0.10   &  0.26&$\pm$&0.13  & $\sim$0.26&  -  &0.39             &  0.33&$\pm$&0.10   \\
$R_\mathrm{sec}/R_\mathrm{L\mathrm{sec}}$  &    1.0&$\pm$&0.6    &   0.3&$\pm$&0.1   &   0.6&$\pm$&0.2   &    0 6&$\pm$&0.5    &   0.7&$\pm$&0.5   & $\sim$0.4&  -  &0.6               &   0.3&$\pm$&0.1    \\
$i [^\circ]$                               &      8&$\pm$& 1     &    51&$\pm$& 7    &    53&$\pm$& 8    &     56&$\pm$&11     &    70&$\pm$&22    & $\ga50^{\circ}$&,&$\la75^{\circ}$   &    53&$\pm$&18   \\
\Teff [k]                                  &  16100&$\pm$&4400   & 16600&$\pm$&1600  &  \multicolumn{3}{c}{---}  &  10400&$\pm$& 200   & 14100&$\pm$& 500  &   35800&$\pm$&300           & 13300&$\pm$&2800 \\
$\log g$                                   &    9.0&$\pm$&0.7    &   8.5&$\pm$&0.3   &  \multicolumn{3}{c}{---}  &    8.0&$\pm$&0.2    &  7.6&$\pm$&0.1    &    7.40&$\pm$&0.05          &  8.4&$\pm$&0.7   \\
\hline\noalign{\smallskip}
\end{tabular}
\end{center}
\end{flushleft}  
\end{table*}  

In    addition,    knowing    that   $M_\mathrm{sec}/M_\mathrm{wd}    =
K_\mathrm{wd}/K_\mathrm{sec}=q$,  we can  also  estimate the  expected
orbital    velocity    of     the    white    dwarf    $K_\mathrm{wd}$
(Table\,\ref{t-param}). The predicted $K_\mathrm{wd}$ amplitudes could
easily  be measured,  e.g.  using  \textit{HST/COS}  observations, and
such  measurements  would be  very  valuable  to  improve the  overall
constraints on the system parameters  of these PCEBs.  Finally, we can
estimate the orbital separations and Roche lobe radii of the secondary
stars from Kepler's third law and Eggleton's (\citeyear{eggleton83-1})
expression
\begin{equation}
R_\mathrm{L\mathrm{sec}}  = \frac{a\,0.49\,q^{2/3}}
{0.6\,q^{2/3} + \ln(1+q^{1/3})}
\label{e-rl}
\end{equation}

\section{Notes on individual systems}
\label{s-notes}

\subsection{SDSS\,J0052--0053, a detached CV in the period gap?}

SDSS\,J0052--0053 has the shortest orbital period, the smallest radial
velocity   amplitude,   and   lowest   inclination   in   our   sample
(Fig.\,\ref{f-scargle},  Table\,~\ref{t-param}).   Another  intriguing
feature of  SDSS\,J0052--0053 is that the Roche  lobe secondary radius
and  the  secondary  star   radius  overlap  within  the  errors  (see
Table\,\ref{t-param}).  The  SDSS and Magellan  spectra certainly rule
out ongoing mass transfer,  i.e. that SDSS\,J0052--0053 is a disguised
cataclysmic variable  (CV). Two possible scenarios could  apply to the
system. SDSS\,J0052--0053 could be either be a pre-CV that is close to
develop in a semi-detached configuration, or it could be a detached CV
in the  2--3\,h period  gap.  Standard evolution  models based  on the
disrupted magnetic  braking scenario \citep[e.g.][]{rappaportetal83-1,
kolb93-1, howelletal01-1} predict that  CVs stop mass transfer and the
secondary star  shrinks below its  Roche lobe radius once  they evolve
down  to $\Porb\simeq3$\,h.  Subsequently,  these detached  CVs evolve
through the period gap, until  the companion star fills its Roche lobe
again at  $\Porb\simeq2$\,h. Population models based  on the disrupted
magnetic braking hypothesis predict that  the ratio of detached CVs to
pre-CVs (with appropriate companion  star masses to (re-)initiate mass
transfer at $\Porb\simeq2$\,h) should be $\ga4$ \citep{davisetal08-1},
hence a  substantial number of such  systems is expected  to exist. So
far,  only  one  other   system  with  similar  properties  is  known,
HS\,2237+8154 \citep{gaensickeetal04-1}.

\subsection{SDSS\,J1724+5620, a PCEB with a strong heating effect}
\label{s-sdss1724}

SDSS\,J1724+5620 contains the hottest  white dwarf among our sample of
PCEBs,  which affects the  determination of  its system  parameters in
several ways. Firstly, irradiation heating the inner hemisphere of the
companion will  cause it  to appear of  earlier spectral type  than an
unheated star of same mass. This effect is observed in the analysis of
the three  SDSS spectra available  in DR6. The  spectral decomposition
types  the companion  star as  M3-4,  and combining  the flux  scaling
factor of  the M star  template and the spectral  type-radius relation
\citep[see][for   details]{rebassa-mansergasetal07-1}   indicates   an
average distance  of $d_\mathrm{sec}=633\pm144$\,pc.  This  is roughly
twice the  distance implied from the  model fit to  the residual white
dwarf  spectrum,  $d_\mathrm{wd}=354\pm15$\,pc,  indicating  that  the
spectral type of the companion  star determined from the SDSS spectrum
is too  early for its  actual mass and  radius, as expected  for being
heated by the  white dwarf. Fixing the spectral  type of the companion
to     M5,      the     spectral     decomposition      results     in
$d_\mathrm{sec}=330\pm150$\,pc, consistent with  the distance based on
the white dwarf fit.

An  additional  complication  is  that  the  radial  velocity  of  the
companion star was  measured from the H$\alpha$ emission  line, as the
\Ion{Na}{I} absorption  doublet was too weak in  the individual 900\,s
SDSS spectra.  The equivalent width  of the H$\alpha$ emission shows a
noticeable  variation as  a  function of  binary  phase, with  maximum
equivalent with near  the maximum in the light  curve, indicating that
the H$\alpha$ emission is concentrated  on the inner hemisphere of the
companion   star  (Fig.\,\ref{f-rvphot1724}).    A  phase   offset  of
$0.046\pm0.008$ is  observed between the photometry  and the H$\alpha$
equivalent  width,  which  we  attribute  to  systematic  problems  in
measuring  the  equivalent width  of  the  H$\alpha$  embedded in  the
photospheric  absorption line  from the  white dwarf,  given  the poor
quality  of  the individual  SDSS  subspectra.   As  a consequence  of
H$\alpha$  predominantly originating  on the  inner hemisphere  of the
companion star,  the observed $K_\mathrm{sec}$ is  an underestimate of
the  true  radial  velocity  amplitude  \citep[e.g.][]{wade+horne88-1,
oroszetal99-1,  vennesetal99-2,   aungwerojwitetal07-1}.   The  radial
velocity  amplitude of  the secondary  star's  centre of  mass can  be
written as \citep{wade+horne88-1}:
\begin{equation}
K_\mathrm{sec,corr} = \frac{K_\mathrm{sec}}{1-(1+q)(\Delta R/a)},
\end{equation}
where $\Delta R$  is the displacement of the centre  of light from the
centre of mass.   $\Delta R = 0$ implies that the  centre of light and
the centre of mass coincide,  whilst $\Delta R = R_\mathrm{sec}$ gives
the maximum possible displacement.  If one assumes that the irradiated
emission  on the  secondary is  distributed uniformly  over  the inner
hemisphere, and  that the contribution  of the irradiation is  zero on
its  un-irradiated  face,  then  $\Delta R  =  (4/3\pi)R_\mathrm{sec}$
\citep{wade+horne88-1, oroszetal99-1,  vennesetal99-2}.  Assuming that
the spectral  type is  in the range  M3--5, different  combinations of
secondary   mass   and   radius,   and  radial   velocity   amplitude,
K$_\mathrm{sec,corr}$,  can then constrain  the orbital  parameters of
SDSS\,J1724+5620 (Table\,\ref{t-param}).

\begin{table*}
\caption{\label{t-evol}   Parameters  of   the   PCEBs  derived   from
  calculating     their     post-CE     evolution     according     to
  \citet{schreiber+gaensicke03-1}.        $t_{\mathrm{cool}}$      and
  $t_{\mathrm{sd}}$ are  the cooling age and the  predicted time until
  the     system     enters     the     semi-detached     CV     phase
  respectively.  $P_{\mathrm{sd}}$  denoted  the zero-age  CV  orbital
  period while $P_{\mathrm{CE}}$ is  the orbital period the system had
  when it left  the CE phase.  As discussed in  the text and displayed
  in Fig\,\ref{f-evol}, the predicted evolution depends sensitively on
  the mass and the radius of  the secondary star, in particular as the
  errors  of all  estimated secondary  masses overlap  with  the value
  assumed  for the  fully convective  mass limit.   We  therefore give
  values for the both scenarios.}  \setlength{\tabcolsep}{0.5ex}
\begin{tabular}{lcc@{\hspace*{3ex}}ccc@{\hspace*{3ex}}ccc}
\hline
\hline
&&& \multicolumn{3}{c}{CMB} & \multicolumn{3}{c}{GR} \\
\multicolumn{1}{c} {Name} & 
\Porb & 
$t_{\mathrm{cool}}$ & $P_{\mathrm{sd}}$ & $P_{\mathrm{CE}}$ & $t_{\mathrm{sd}}$ &$P_{\mathrm{sd}}$ & $P_{\mathrm{CE}}$ & $t_{\mathrm{sd}}$ \\
SDSS\,J & [days] & [years] & [days] & [days] & [years] & [days] & [days] & [years] \\
\hline
0052--0053 & 0.114 & 4.2\e{8}& 0.125 & 0.46 & 0.0     & 0.114 & 0.145 & 0.0\\
0246+0041  & 0.726 & 3.1\e{8}& 0.146 & 0.78 & 1.1\e{9}& 0.113 &  0.73 & 8.1\e{10}\\
0314--0111 & 0.263 & ---     & 0.12  & ---  & 3.7\e{7}&  0.11 &   --- & 6.2\e{9}\\
1151--0007 & 0.142 & 5.0\e{8}& 0.12  & 0.55 & 2.7\e{6}&  0.07 &  0.15 & 1.8\e{9}\\
1529+0020  & 0.165 & 1.3\e{8}& 0.11  & 0.41 & 4.7\e{6}&  0.10 &  0.17 & 2.5\e{9}\\
1724+5620  & 0.333 & 3.2\e{6}& 0.12  & 0.34 & 6.1\e{7}&  0.11 &  0.33 & 1.7\e{10}\\
2339--0020 & 0.656 & 5.1\e{8}& 0.12  & 0.74 & 1.0\e{9}&  0.11 &  0.66 & 6.5\e{10}\\
\hline\noalign{\smallskip}
\end{tabular}
\end{table*}

\subsection{SDSS\,J0309-0101, SDSS\,J1138-0011 and SDSS\,J2241+0027:
wide WDMS binaries?}
\label{s-nonPCEB}

\begin{figure}
\includegraphics[width=0.143\textwidth, angle=-90]{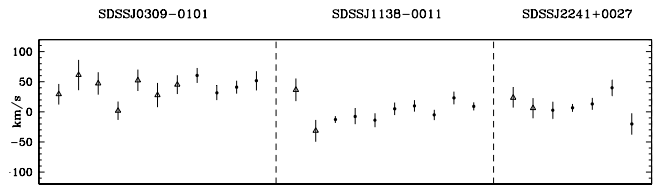}
\caption{\label{f-rvWDMS}  SDSS  radial  velocities (triangles)  along
with the  \Ion{Na}{I} radial velocities  measured in this  work (solid
dots) of SDSS\,J0309-0101,  SDSS\,J1138-0011 and SDSS\,J2241+0027. The
data at  hand suggest that this  three systems are  wide WDMS binaries
instead of PCEBs.}
\end{figure}

SDSS\,J0309-0101,  SDSS\,J1138-0011 and SDSS\,J2241+0027  were flagged
as PCEB candidates by \citet{rebassa-mansergasetal07-1} on the base of
a $3\sigma$ radial velocity  variation in between their different SDSS
spectra, as  measured from either  the H$\alpha$ emission line  or the
\Ion{Na}{I} absorption  doublet from the companion  star. However, the
additional  intermediate-resolution  spectra  taken  for  these  three
objects  (Table~\ref{t-logobs})  do  not  show  a  significant  radial
velocity variation  (Table\,\ref{t-rvs}, Fig.\,\ref{f-rvWDMS}).  It is
therefore important to review the  criterion that we used in our first
paper to identify  PCEBs from repeated SDSS spectroscopy  in the light
of two subtle changes.

On the  one hand, we have  slightly modified the procedure  to fit the
\Ion{Na}{I} absorption doublet,  as outlined in Sect.\,\ref{s-rvs}. By
comparing    the     \Ion{Na}{I}    DR5    radial     velocities    in
\citet{rebassa-mansergasetal07-1}  with those  obtained  with the  new
procedure  for the  same spectra,  we  find an  average difference  of
5\,\kms\, with a  maximum difference of 10.5\,\kms.  In  all cases the
measurements agree within the errors.

On    the   other    hand,   we    were   using    DR5    spectra   in
\citet{rebassa-mansergasetal07-1}, but  the analysis carried  out here
was  done using  DR6 spectra,  which were  processed with  a different
reduction  pipeline  \citep{adelman-mccarthyetal08-1}.   A  comparison
between  the   DR5  and  DR6   radial  velocity  values   obtained  in
\citet{rebassa-mansergasetal07-1} and in  this work respectively (both
measured following our new procedure) provides in this case an average
difference of 6.5\,\kms, with a maximum of 22\,\kms. Again, the radial
velocity  measurements   agree,  with   the  exception  of   a  single
spectrum, within the errors.

The conclusion from  comparing our two methods, and  the two SDSS data
releases,  is that the  radial velocity  measurements obtained  are in
general consistent  within their errors.  However, the  offsets can be
sufficient to move  a given system either way  across our criterion to
identify PCEB  candidates, being defined a  3$\sigma$ radial variation
between their SDSS spectra. This  is specifically the case for systems
with either low-amplitude radial velocity variations, or faint systems
with  noisy spectra.   An  additional  note concerns  the  use of  the
H$\alpha$   line   as    probe   for   radial   velocity   variations.
\citet{rebassa-mansergasetal07-1}  identified  SDSS\,J2241+0027  as  a
PCEB candidate on the basis of  a large change in the H$\alpha$ radial
velocity  between  the  two   available  SDSS  spectra.  However,  the
velocities  obtained  from  the  \Ion{Na}{I} doublet  did  not  differ
significantly.  Inspecting  the spectra again confirms  the results of
\citet{rebassa-mansergasetal07-1}.  This  suggests that the absorption
lines from  the secondary star are  a more robust probe  of its radial
velocity.

A somewhat speculative explanation for the shifts found from H$\alpha$
radial  velocity  measurements  is  that  the main  sequence  star  is
relatively rapidly rotating, and that the H$\alpha$ emission is patchy
over its  surface.  A  nice example  of a this  effect is  the rapidly
rotating  ($P=0.459$\,d)   active  M-dwarf  EY\,Dra,   which  displays
H$\alpha$ radial velocity variations  with a peak-to-peak amplitude of
$\sim100\,\kms$  \citep{eibe98-1}. In order  to explain  the H$\alpha$
radial   velocity   shifts   of   a   few   $10\,\kms$   observed   in
e.g. SDSS\,J2241+0027,  the companion star  should be rotating  with a
period  of  $\sim1$\,d.  According to  \citet{cardini+cassatella07-1},
low-mass stars such as the companions  in our PCEBs that are a few Gyr
old are  expected to have rotational  periods of the order  of tens of
days, which  is far too long  to cause a  significant H$\alpha$ radial
velocity  variation at the  spectral resolution  of the  SDSS spectra.
However,  \citeauthor{cardini+cassatella07-1}'s  study  was  based  on
single stars, and the rotation  rates of the companion stars in (wide)
WDMS binaries is not well established.

\section{PCEB evolution}
\label{s-evol}

\begin{figure}
\includegraphics[width=8.3cm]{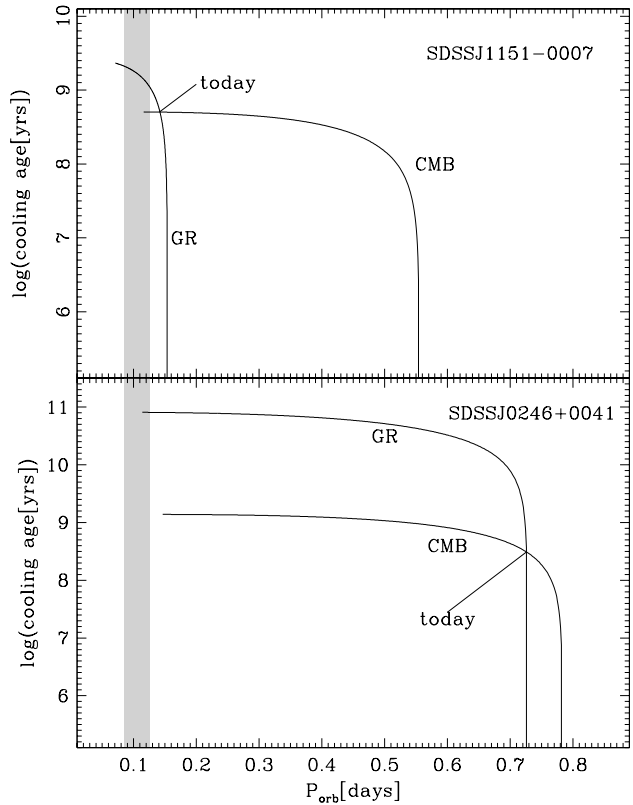} 
\caption{\label{f-evol}  The predicted  evolution of  two  PCEBs.  The
  orbital period  gap observed in  the orbital period  distribution of
  CVs  is shaded.  As  the secondary  masses of  SDSS\,J1151--0007 and
  SDSS\,J0246+0041  are close  to  the fully  convective boundary,  we
  calculated the  evolution assuming classical  magnetic braking (CMB)
  and  assuming  only gravitational  radiation  (GR).  Obviously,  the
  calculated   orbital  periods   at   the  end   of   the  CE   phase
  ($P_{\mathrm{CE}}$), the  evolutionary time scale,  and the expected
  zero-age    CV    orbital    period    ($P_{\mathrm{sd}}$)    differ
  significantly. }
\end{figure}

Following \citet{schreiber+gaensicke03-1} we determine the cooling age
and the  future evolution  of the new  PCEBs. We assume  classical and
disrupted  magnetic braking  according to  the standard  theory  of CV
evolution  \citep[e.g.][]{verbunt+zwaan81-1}.   We  used  the  cooling
tracks of \citet{wood95-1} and find that most of the PCEBs left the CE
about  $1-5\times\,10^8$  years ago,  the  only exception  being
SDSS\,J1724+5620  which appears to  be much  younger.  The  time still
needed to enter the semi-detached CV configuration is shorter than the
Hubble time  for all the  systems, qualifying them  as representatives
for  the progenitors  of the  current CV  population.  The  numbers we
obtained   from   the   theoretical   analysis   are   summarised   in
Table\,\ref{t-evol}. The estimated secondary masses are generally very
close  to   the  fully  convective  boundary  (for   which  we  assume
$M_\mathrm{cc}=0.3M_{\odot}$,   see    however   the   discussion   by
\citealt{mullan+macdonald01-1}  and \citealt{chabrieretal07-1}  on how
this  mass  limit  may  be  affected by  magnetic  activity)  and  the
uncertainties involved  in the secondary mass  determination are quite
large.   The predicted evolution  of all  systems should  therefore be
considered  highly  uncertain as  it  is  not  clear whether  magnetic
braking applies or not.   We therefore give in Table\,\ref{t-evol} the
values obtained for both scenarios.  The effect of not knowing whether
magnetic  braking  or  only  gravitational radiation  will  drive  the
evolution of  the systems is illustrated  in Fig.\,\ref{f-evol}, which
shows  the  expected  post  CE  evolution  for  SDSS\,J1152--0007  and
SDSS\,J0246+0041 for both cases.

Inspecting  Table\,\ref{t-evol},  the  case  of  SDSS\,J0052--0053  is
particularly   interesting.   The   estimated   secondary  radius   is
consistent with its Roche-lobe radius and the system is supposed to be
very  close to  the  onset of  mass  transfer, which  is reflected  by
$t_\mathrm{sd}=0$ within  the accuracy of  our calculations.  Clearly,
SDSS\,J0052--0053 may be a detached CV  in the period gap, or a pre-CV
that  has  almost  completed  its  PCEB  lifetime.   To  estimate  the
probability for  SDSS\,J0052--0053 being either  a period-gap CV  or a
PCEB close  to entering for the  first time a  semi-detached state, we
assume a steady  state binary population. In that  case, the number of
detached CVs in  the gap should be roughly equal to  the number of CVs
above the gap \citep[][]{kolb93-1}.   In addition, the number of PCEBs
in the gap  should be roughly equal to the number  of accreting CVs in
the gap.  According to CV population studies \citep[][]{kolb93-1}, the
number of detached  CVs in the gap is  approximately five times higher
than  the  number   of  CVs  in  the  gap   and  the  probability  for
SDSS\,J0052--0053 being a detached CV in the gap rather than a PCEB is
$\sim\,80\%$.  This  result is a  simple consequence of the  fact that
all CVs  with donor stars  earlier than $\sim$M3 will  become detached
CVs in the gap while only a rather small fraction of PCEBs, i.e. those
with secondary  stars of spectral type  $\sim$M3.5--M4.5, produces CVs
starting mass  transfer in  the period gap.  This is obviously  only a
rough estimate.  A detailed population  study by \citet{davisetal08-1}
confirms  that the  ratio of  detached CVs  to pre-CVs  within  the CV
period gap  is $\sim4-13$, depending  on different assumptions  of the
common envelope ejection efficiency, initial mass ratio distributions,
and magnetic braking laws.

\section{Discussion}
\label{s-why}

As  outlined in  the introduction,  despite some  recent  progress our
understanding of  the CE phase  is still very limited.   The classical
$\alpha$ prescription  \citep{webbink84-1} that  is based on  a simple
energy equation seems  to be unable to explain  the observed existence
of double  white dwarfs with  stellar components of  comparable masses
and    rather    long    orbital    periods    \citep[see][for    more
details]{webbink07-1}.         This       disagreement       motivated
\citet{nelemansetal00-1} to develop  an alternative prescription based
on  the  conservation  of  angular momentum,  the  so-called  $\gamma$
algorithm.  Binary  population synthesis  codes based on  the $\alpha$
prescription as well as  those assuming the $\gamma$ algorithm predict
the existence  of a significant  number of PCEBs with  orbital periods
longer    than    a    day.     Using    the    $\alpha$    mechanism,
\citet{willems+kolb04-1}  calculated the expected  period distribution
of the  present-day WDMS binaries  in the Galaxy  at the start  of the
WDMS binary  phase.  Their Fig.\,10  clearly shows that  the predicted
PCEB  distribution peaks  around $\Porb\sim\,1$\,day,  but also  has a
long tail of systems with up to $\sim100$\,d.  The $\gamma$ algorithm,
on the  other hand, predicts an  increase of the number  of PCEBs with
increasing  orbital period  up to  $\Porb\gappr\,100$ days,  i.e.  the
$\gamma$ prescription predicts the existence of even more long orbital
period                      ($\Porb>1$\,day)                     PCEBs
\citep[see][]{nelemans+tout05-1,maxtedetal07-2}.

\begin{figure}
\includegraphics[angle=270,width=8.3cm]{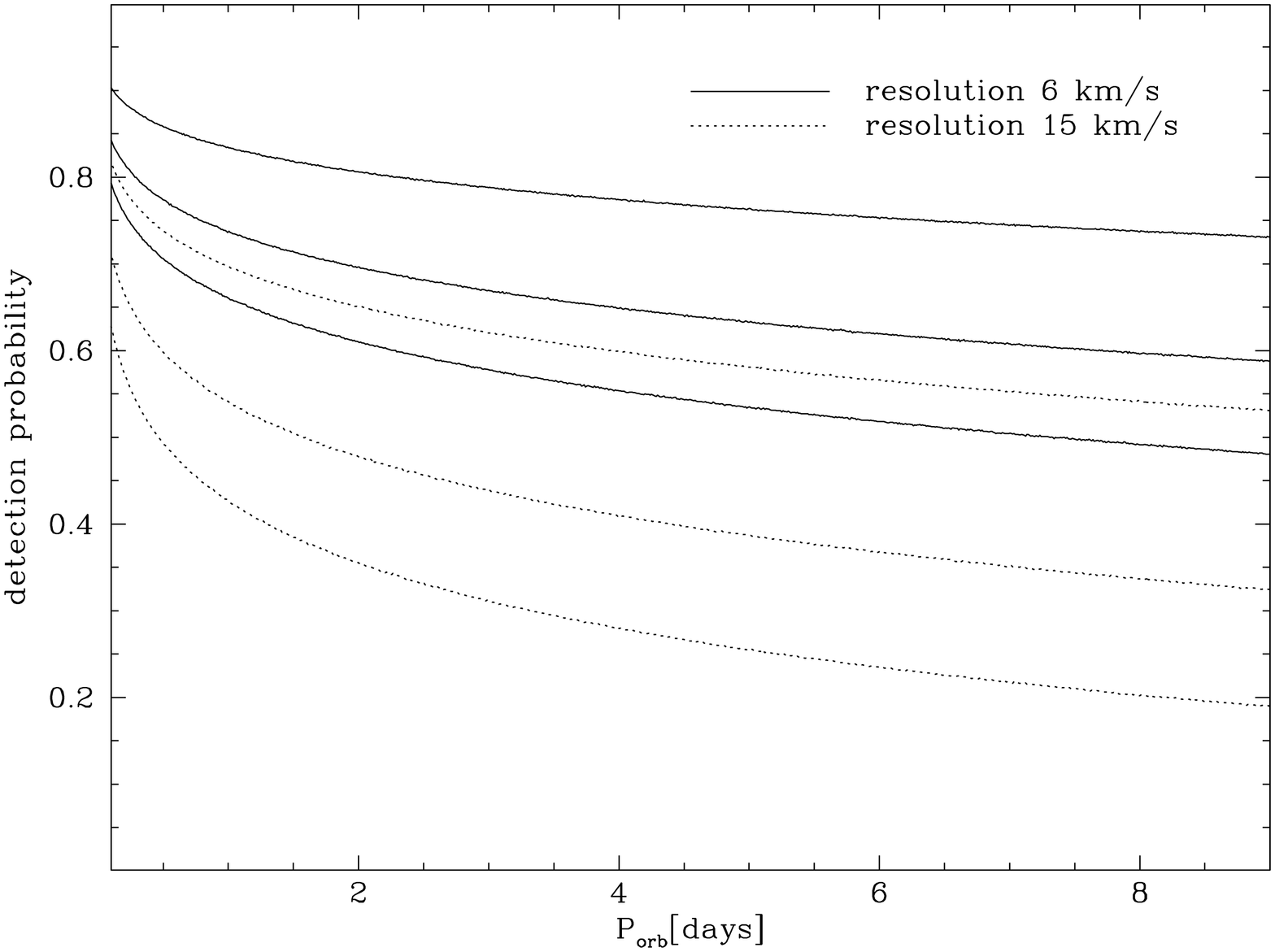} 
\caption{\label{f-dprob}  Monte-Carlo  simulations  of  the  detection
  probability  of  significant  radial  velocity  variations  assuming
  measurement  accuracy  of  6\,\kms,  corresponding to  our  previous
  VLT/FORS   observations   \citep{schreiberetal08-1}   $15$km/s,   as
  appropriate for  the SDSS spectra \citep{rebassa-mansergasetal07-1}.
  The three lines correspond to 1, 2, or 3$\sigma$ significance of the
  radial  velocity variation.  Clearly,  even the  PCEB identification
  based on  multiple SDSS  spectra should be  sensitive to  PCEBs with
  orbital periods of $\sim\,1-10$\,days.}
\end{figure}

We have  measured a  total number of  nine orbital periods  along this
work      and      the      recently      published      paper      by
\citet{schreiberetal08-1}. All  of them have short  orbital periods of
less  than a  day.   Radial  velocity variations  are  much easier  to
identify in short orbital period systems, so our sample is expected to
be biased  towards shorter orbital  periods.  The crucial  question is
whether  our   observational  finding,  the  paucity   of  PCEBs  with
$\Porb>1$\,d,  is whether  the  result  of that  bias,  or whether  it
reflects an intrinsic  feature of the PCEB population.  To answer this
question we  performed the following analysis.  First,  we carried out
Monte-Carlo    simulations    similar    to   those    presented    in
\citet[][]{schreiberetal08-1}
\footnote{In     \citet[][their     Fig.\,8]{schreiberetal08-1}     we
underestimated  the  detection  probability.  Therefore,  the  correct
detection  probability  for a  resolution  of  $6$\,km/s  as given  in
Fig.\,\ref{f-dprob}  is  higher  than  our previous  estimates.}   but
assuming a resolution of  $15$km/s, corresponding to the typical error
in the  radial velocity measurements  from SDSS spectra.   Clearly, as
shown in  Fig.\,\ref{f-dprob}, the  bias towards short  orbital period
systems is larger due to the lower resolution of the SDSS spectra and,
as a consequence, radial  velocity variations in multiple SDSS spectra
of  systems  with  $\Porb>10$\,days  are  more  difficult  to  detect.
However,   the  probability  to   detect  $3\sigma$   radial  velocity
variations   for   systems   in    the   orbital   period   range   of
$\Porb\sim1-10$\,days  still is $\sim20-40\%$.   Second, we  assumed a
uniform    orbital   period    distribution    and   integrated    the
$3\sigma$-detection probability (lowest  dotted line) for systems with
$\Porb  \leq 1$\,day  and those  with $10$  days  $>\Porb>1$\,day.  In
other  words, we  assume  that  there is  no  decrease for  $10$\,days
$\Porb>1$\,day and calculate the  total detection probability for $10$
days $>\Porb>1$\,day  and $\Porb \leq 1$  day.  We find  that if there
was indeed no  decrease, the fraction of PCEBs  with $\Porb>1$\,day in
our  sample  should  be  $\sim84\%$  and  the  number  of  PCEBs  with
$\Porb>1$\,day  among our  nine systems  should  be $\sim\,7.6\pm2.8$.
The result  of our observations,  i.e.  no system  with $\Porb>1$\,day
among  nine PCEBs  disagrees with  the hypothesis  (i.e.  there  is no
decrease) by  $2.7\sigma$.  This indicates  that the measured  lack of
PCEBs with $\Porb>1$\,day  might indeed be a feature  of the intrinsic
population of PCEBs.

\begin{figure}
\includegraphics[width=0.26\textwidth, angle=-90]{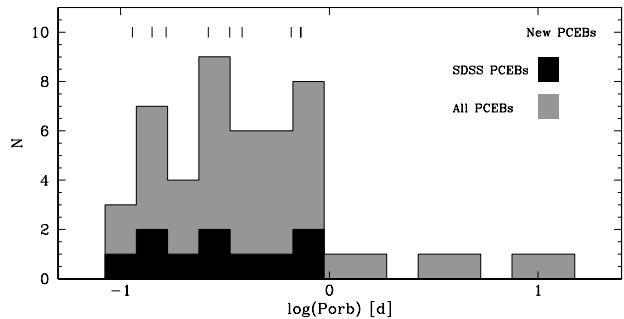}
\caption{\label{f-histo} The period distribution  of 41 PCEBs from the
\citet{ritter+kolb03-1} (V7.9) catalogue containing a white dwarf plus
M or K  companion star (gray). The 7 systems studied  in this work and
the      recent      PCEB     discovery      SDSS\,J143547.87+373338.5
\citep{steinfadtetal08-1}  are also  included.   In black  is shown  a
subsample  of SDSS  PCEBs obtained  from RV  variation  studies.  Both
subsamples  illustrate a  clear  lack of  systems  at orbital  periods
greater than 1 day.  The 7 new  systems from this work and the the two
systems in  \citet{schreiberetal08-1} are  indicated as tick  marks in
the top of the figure.}
\end{figure}

How do our results for PCEBs  from SDSS compare with the sample of all
known PCEBs?   Figure\,\ref{f-histo} shows in gray  the orbital period
distribution of all 41 known PCEBs  consisting of a M or K-star plus a
white  dwarf from  the latest  version of  the \citet{ritter+kolb03-1}
catalogue  (V7.9), including  also  the seven  new  periods from  this
paper, and the new  recent discovery by \citet{steinfadtetal08-1}.  As
we are interested in analysing  the bias of the sample towards shorter
orbital  periods   in  the   context  of  detection   probability,  we
superimpose in  black a subsample  of ten SDSS PCEBs,  identified from
radial  velocity snapshots  of  SDSS WDMS,  representing a  homogenous
selection  mechanism   (seven  from   the  current  paper,   two  from
\citealt{schreiberetal08-1},   one   [SDSS\,J112909.50+663704.4]  from
\citealt{raymondetal03-1}).  So  far, the period  distributions of the
two subsets seem very similar, with  a steep decrease of the number of
systems at $\Porb\sim1$\,day.  The total sample of PCEBs contains only
six systems  with orbital periods  larger than one day  ($\sim$15\% of
the entire sample).  This suggests  that the number of PCEBs decreases
for  $\Porb>1$\,day,  implying  that  the $\gamma$  algorithm  in  its
present form might not be an adequate description of the CE phase, and
that the common envelope efficiency in the energy equation used in the
$\alpha$  prescription  is  perhaps  smaller than  assumed  previously
(lower efficiencies lead to stronger shrinkage of binary separations).
However, one should bear in  mind that the previously known PCEBs have
been  discovered by  various  methods,  and therefore  do  not form  a
representative                  sample                 \citep[see][for
details]{schreiber+gaensicke03-1}. The  sample of PCEBs  selected in a
homogeneous way from  the SDSS is still small,  and the current sample
of  WDMS  binaries from  SDSS  involves  some  selection effects,  too
\citep{schreiberetal07-1},  impeding  a  definite conclusion  at  this
point.  However,  work is underway  to enlarge the parameter  space of
the  SDSS WDMS binaries  in terms  of ages  and secondary  star masses
\citep{schreiberetal07-1},  to  increase  the  number  of  PCEBs  with
orbital  period measurements,  and  to model  the remaining  selection
effects, paving the  way for a more quantitative  assessment of CE and
post-CE evolution.

\section{Conclusions}

We have presented a study of 11 PCEB previously identified candidates,
and measured  the orbital periods  of six and  one of them  from their
\Ion{Na}{I}  doublet radial velocity  variations and  its differential
photometry,  respectively.   Combining  the $K_\mathrm{sec}$  velocity
amplitudes with  the results from  spectroscopic decomposition/fitting
of their  SDSS spectra,  we constrained the  binary parameters  of the
seven systems for which we could determine orbital periods.  No radial
velocity   variations  were  detected   for  three   PCEB  candidates,
suggesting  that they  may be  wide binaries.   We revisited  the PCEB
candidate selection of \citet{rebassa-mansergasetal07-1}, and conclude
that  candidates with  low amplitude  velocity variations,  noisy SDSS
spectra, and radial velocity shifts that only show up in the H$\alpha$
emission  line definitely need  additional follow-up  spectroscopy for
the confirmation/rejection of their PCEB nature.  Finally, we have had
a first look at the  period distribution of PCEBs from SDSS identified
from radial velocity  snapshots of SDSS WDMS binaries,  and noted that
none  of the  ten  systems published  so  far have  an orbital  period
$>1$\,d.   Using a  Monte-Carlo simulation,  we demonstrated  that our
method  of finding  PCEBs should  be efficient  also for  systems with
periods $>1$\,d,  and that, subject  to small
number statistics, it appears  that the PCEB period distribution peaks
at $\Porb<1$\,d.  This is in agreement with the period distribution of
the  previously known PCEBs,  which represents  a heterogenous  mix of
systems identified by various  different methods. In contrast, current
binary population models predict a  large number of PCEBs with orbital
periods $>1$\,day  in clear  disagreement with the  currently observed
sample. Additional effort is needed  to improve the size of the sample
of known PCEBs,  as well as to fully model  its selection effects, but
with more than  1000 SDSS WDMS binaries to draw  from, the outlook for
more quantitative tests of CE evolution seem promising.

\section*{Acknowledgements.}
ARM was supported by  a STFC-IAC studentship. MRS acknowledges support
from FONDECYT (grant  1061199), DIPUV (project 35), and  the Center of
Astrophysics in Valparaiso (CAV). ANGM, AST, RSC, JV were supported by
the DLR  under grant  50OR0404.  AST acknowledges  support by  the DFG
grant  Schw536/20-1.  We  thank  the  referee,  John  Thorstensen  for
providing comments that helped improving the paper.

Based  on  observations collected  at  the  European Organisation  for
Astronomical  Research  in   the  Southern  Hemisphere,  Chile,  under
programmes  079.D-0531  and 080.D-0407;  at  the Centro  Astron\'omico
Hispano  Alem\'an at Calar  Alto, operated  jointly by  the Max-Planck
Institut  f\"ur  Astronomie  and  the Instituto  de  Astrof\'isica  de
Andaluc\'ia; at the  WHT, which is operated on the  island of La Palma
by the Isaac Newton Group in the Spanish Observatorio del Roque de los
Muchachos of  the Instituto  de Astrof´isica de  Canarias; and  at the
1.2-m  telescope,  located at  KryoneriKorinthias,  and  owned by  the
National  Observatory  of Athens,  Greece.  This  paper includes  data
gathered  with the 6.5m  Magellan Telescopes  located at  Las Campanas
Observatory, Chile.

\label{lastpage}

\end{document}